\documentclass[useAMS,usenatbib,fleqn]{mn2e}

\usepackage[draft]{hyperref}
\usepackage{amssymb}
\usepackage{amsmath}
\usepackage{graphicx}

\usepackage{nicefrac}
\usepackage{bigints}
\usepackage{color}
 
\title[SNe Ia from violent mergers: polarisation signatures]{Type Ia supernovae from violent mergers of carbon-oxygen white dwarfs: polarisation signatures}
\author[M. Bulla et al.]{M.~Bulla,$^1$\thanks{E-mail: mbulla01@qub.ac.uk} S.~A.~Sim,$^{1,2}$ R.~Pakmor,$^3$ M.~Kromer,$^4$ S.~Taubenberger,$^{5,6}$  \newauthor F.~K.~R\"{o}pke,$^{3}$ W.~Hillebrandt$^6$ and I.~R. Seitenzahl$^{2,7}$\\
$^1$Astrophysics Research Centre, School of Mathematics and Physics, Queen's University Belfast, Belfast BT7 1NN, UK\\
$^2$ARC Centre of Excellence for All-sky Astrophysics (CAASTRO), Australian National University, Canberra, ACT 2611, Australia\\
$^3$Heidelberger Institut f\"{u}r Theoretische Studien, Schloss-Wolfsbrunnenweg 35, 69118 Heidelberg, Germany\\
$^4$The Oskar Klein Centre and Department of Astronomy, Stockholm University, AlbaNova, SE-106 91 Stockholm, Sweden\\
$^5$European Southern Observatory, Karl-Schwarzschild-Str. 2, D-85748 Garching, Germany\\
$^6$Max-Planck-Institut f\"{u}r Astrophysik, Karl-Schwarzschild-Str. 1, D-85748 Garching bei M\"{u}nchen, Germany\\
$^7$Research School of Astronomy and Astrophysics, Mount Stromlo Observatory, Cotter Road, Weston Creek, ACT 2611, Australia}
\date{Accepted 2015 October 14. Received 2015 October 09; in original form 2015 September 04}

\newcommand{\newCommandName}{pdf}  

\begin{document}

\maketitle 

\begin{abstract}
The violent merger of two carbon-oxygen white dwarfs has been proposed as a viable progenitor for some Type Ia supernovae. However, it has been argued that the strong ejecta asymmetries produced by this model might be inconsistent with the low degree of polarisation typically observed in Type Ia supernova explosions. Here, we test this claim by carrying out a spectropolarimetric analysis for the model proposed by \citet{pakmor2012} for an explosion triggered during the merger of a 1.1~M$_{\odot}$ and 0.9~M$_{\odot}$ carbon-oxygen white dwarf binary system. Owing to the asymmetries of the ejecta, the polarisation signal varies significantly with viewing angle. We find that polarisation levels for observers in the equatorial plane are modest ($\lesssim$~1~per~cent) and show clear evidence for a dominant axis, as a consequence of the ejecta symmetry about the orbital plane. In contrast, orientations out of the plane are associated with higher degrees of polarisation and departures from a dominant axis. While the particular model studied here gives a good match to highly-polarised events such as SN~2004dt, it has difficulties in reproducing the low polarisation levels commonly observed in normal Type Ia supernovae. Specifically, we find that significant asymmetries in the element distribution result in a wealth of strong polarisation features that are not observed in the majority of currently available spectropolarimetric data of Type Ia supernovae. Future studies will map out the parameter space of the merger scenario to investigate if alternative models can provide better agreement with observations.

%the line polarisation predicted by this model is too strongdegrees of polarisation predicted across spectral features are too strong, pointing toward an (excessive) asymmetry of the element distribution.
\end{abstract}
\begin{keywords}
polarisation -- radiative transfer -- hydrodynamics -- methods: numerical -- supernovae: general 
\end{keywords}

\section{Introduction}

Despite much effort, we still lack a comprehensive picture of the progenitor systems and explosion mechanisms of Type Ia supernovae (SNe Ia). The general consensus is that SNe Ia are thermonuclear explosions of carbon-oxygen white dwarf stars in close binary systems (see e.g. \citealt{roepke2011}; and \citealt{hillebrandt2013}, for reviews), but the exact circumstances leading to these events are unclear. 

\begin{figure*}
\begin{center}
\includegraphics[width=0.48\textwidth]{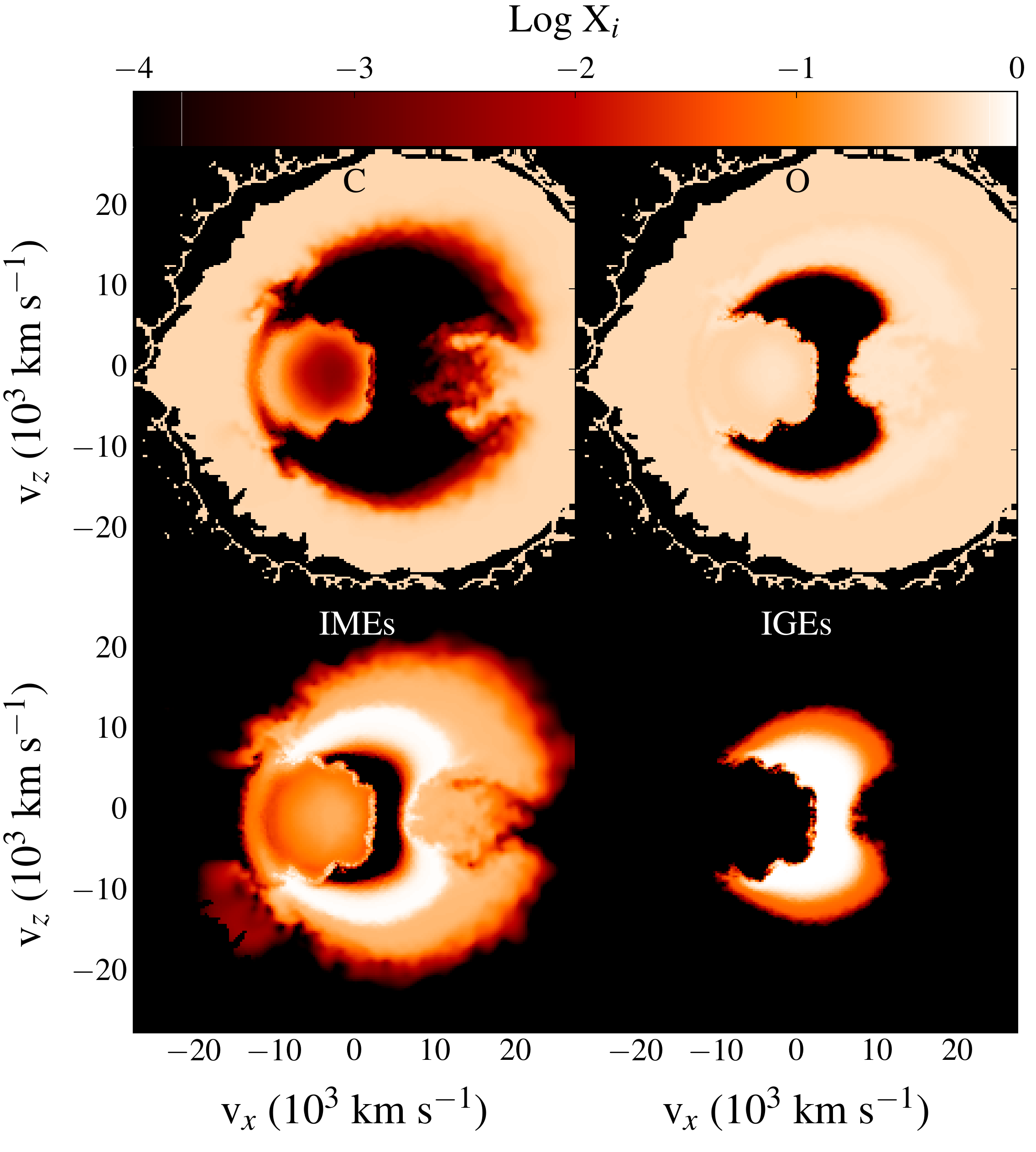}
\includegraphics[width=0.4925\textwidth]{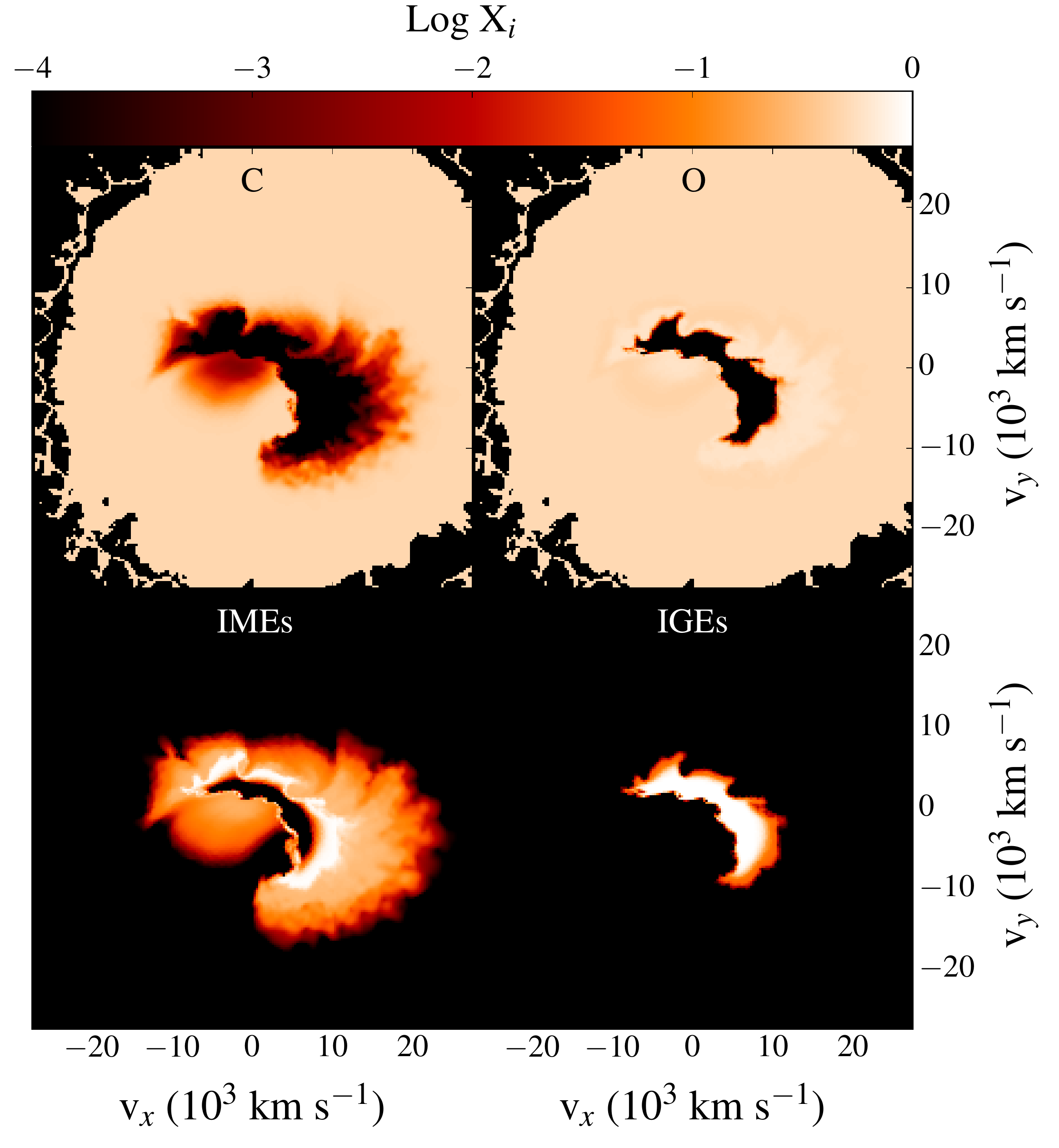}
\caption{Composition of the ejecta for the model of \citet{pakmor2012} 100 s after explosion. The mass fractions of carbon, oxygen, IMEs (from silicon to calcium) and IGEs are mapped to a $200^3$ grid and shown for two slices through the $x-z$ plane (left-hand panels) and the equatorial $x-y$ plane (right-hand panels).} 
%\textcolor{blue}{In the final version use maps with 200$^3$ cubes.}}
\label{ejecta}
\end{center}
\end{figure*}

In the so-called single degenerate scenario \citep{whelan1973}, a carbon-oxygen white dwarf accretes material from a non-degenerate companion and explodes when it approaches the Chandrasekhar limit.
%via either a deflagration-to-detonation transition \citep{khokhlov1991,hoeflich1995,gamezo2005,kasen2009,blondin2013,seitenzahl2013}, a gravitationally confined detonation \citep{plewa2004,jordan2008,jordan2012} or .
However, alternative possibilities have been proposed. Among them, one promising mechanism involves
%the detonation of a layer of helium-rich material on the surface of the accreting white dwarf \citep[the double detonation model;][]{nomoto1980,woosley1980,shen2009,fink2010,woosley2011,moll2013} or 
the merger of two carbon-oxygen white dwarfs \citep{iben1984,webbink1984}. Although the white dwarf merger is able to naturally explain many observed properties of SNe Ia \citep[see e.g.][for a review]{maoz2014}, it is unclear whether the explosion is triggered during the dynamical merger phase or later on. In the latter scenario the primary star is believed to explode inside the merger remnant left behind by the complete disruption of the lighter star \citep{benz1990,vankerkwijk2010,shen2012,zhu2013,dan2014,raskin2014,kashyap2015}. In the former, instead, the violent nature of the accretion process is thought to create conditions for an explosive carbon ignition during the merger \citep{pakmor2010,pakmor2012,pakmor2013,guillochon2010,moll2014}. 

Radiative transfer calculations for violent merger models have shown that the match with spectra and light curves of normal SNe Ia is reasonable for relatively massive white dwarf pairs \citep{pakmor2012,roepke2012,moll2014}. It has also been shown that, for less massive white dwarfs, the model might account for certain sub-luminous SNe Ia like PTF10ops or SN~2010lp \citep{kromer2013}. However, violent merger models have been called into question - particularly for normal SNe Ia - because they produce highly asymmetric ejecta that appear to be in contrast with the low level of polarisation typically observed \citep[see][for a review]{wang2008}. For instance, \citet{maund2013} recently obtained high-quality data for SN~2012fr at four different epochs and found that the polarisation was consistently below 0.1~per~cent. They suggested that such low level of polarisation was inconsistent with the asymmetric $^{56}$Ni distribution predicted by the violent merger model of \citet{pakmor2012}. Comparing predictions of hydrodynamic simulations with observations, however, is not trivial and requires that polarisation calculations are performed for the specific ejecta morphology. 

Here we present an initial spectropolarimetric investigation of the violent merger scenario. Specifically, we focus on the model of \citet{pakmor2012}, which has been used as a benchmark for the violent merger scenario in several previous studies \citep[e.g.][]{roepke2012,summa2013,scalzo2014,seitenzahl2015}. 
%and carry out a polarisation spectral synthesis analysis using the technique recently developed by \citet{bulla2015}. 
We note, however, that the conclusions drawn in this paper may not necessarily be valid in other merger models for which the ejecta morphologies can be substantially different \citep[e.g.][]{pakmor2010,pakmor2013,kromer2013,moll2014}.

%Alternative merger models are known to produce different ejecta morphologies and will be addressed in a separate work.

In Section~\ref{model} we give a brief description of the explosion model of \citet{pakmor2012}, while in Section~\ref{radtransf} we discuss the details of the radiative transfer calculations. We present synthetic observables for the violent merger model in Section~\ref{synth}, together with comparisons to spectropolarimetric data of normal SNe Ia. Finally, we discuss our results and draw conclusions in Section~\ref{conclusions}.

\section{Model}
\label{model}

The binary system studied by \citet{pakmor2012} comprises two white dwarfs with masses of 1.1~M$_{\odot}$ and 0.9~M$_{\odot}$ and initial compositions of 47.5\%~$^{12}$C, 50\%~$^{16}$O and 2.5\%~$^{22}$Ne. In this model, a detonation is assumed to be ignited following the disruption of the secondary star, when material on the surface of the primary reaches sufficiently high densities and temperatures \citep{seitenzahl2009}. The detonation starts to propagate and the energy release from nuclear burning eventually leads to a thermonuclear explosion that unbinds the merging object.

\citet{pakmor2012} followed the evolution of the system until $100$~s after ignition, by which time the ejecta have entered the homologous expansion phase. During the explosion, 0.7~M$_{\odot}$ of iron-group-elements (IGEs) are synthesised, of which the most abundant isotope (0.61~M$_{\odot}$) is radioactive $^{56}$Ni. In addition, 0.5~M$_{\odot}$ of intermediate-mass elements (IMEs) and 0.5~M$_{\odot}$ of oxygen are produced, with 0.15~M$_{\odot}$ of carbon left unburned in low density regions.

Fig. \ref{ejecta} shows two slices through the composition of the ejecta - mapped to a $200^3$ Cartesian grid - $100$~s after explosion. The ejecta structure is highly aspherical, owing to the asymmetry of the merging object at the time of ignition. However, the ejecta are fairly symmetric about the $x-y$ plane in which the inspiral and the merger phases take place before the explosion (i.e. the orbital plane). A distinctive feature is given by the cavity in the IGE distribution. As described by \citet{pakmor2012}, this feature is created as a direct consequence of the different velocities with which the detonation flame propagates inside the merged object. Given that the propagation is faster through high density regions, the primary star burns first and its ashes have expanded considerably by the time the secondary is completely burned. Therefore, the inner parts of the ejecta are dominated by the ashes of the secondary and do not contain IGEs. 

\section{Radiative transfer calculations}
\label{radtransf}

Although radiative transfer calculations for the model introduced in Section \ref{model} have already been presented in \citet{pakmor2012}, here we report new simulations that include polarisation. As in the previous study, we remap the model to a $50^3$ Cartesian grid and perform calculations by using our 3D Monte Carlo radiative transfer code Applied Radiative Transfer In Supernovae \citep[\textsc{artis};][]{sim2007,kromer2009}. Assuming local thermodynamic equilibrium for the first 10 time-steps ($t<2.95$\,d) and a grey approximation for optically thick cells \citep{kromer2009}, we follow the propagation of $N_\text{q}$ Monte Carlo quanta over 111 logarithmic time-steps from 2 to 120 days after explosion. While \citet{pakmor2012} used $N_\text{q}=1\times10^8$ and restricted to a simplified atomic data set (``cd23\_gf-5", see Table 1 of \citealt{kromer2009}), here we utilise $N_\text{q}=2\times10^8$ and the more extended atomic data set described by \citet{gall2012}.

\begin{figure}
\begin{center}
\includegraphics[width=0.45\textwidth,trim=0pt 0pt 0pt -10pt]{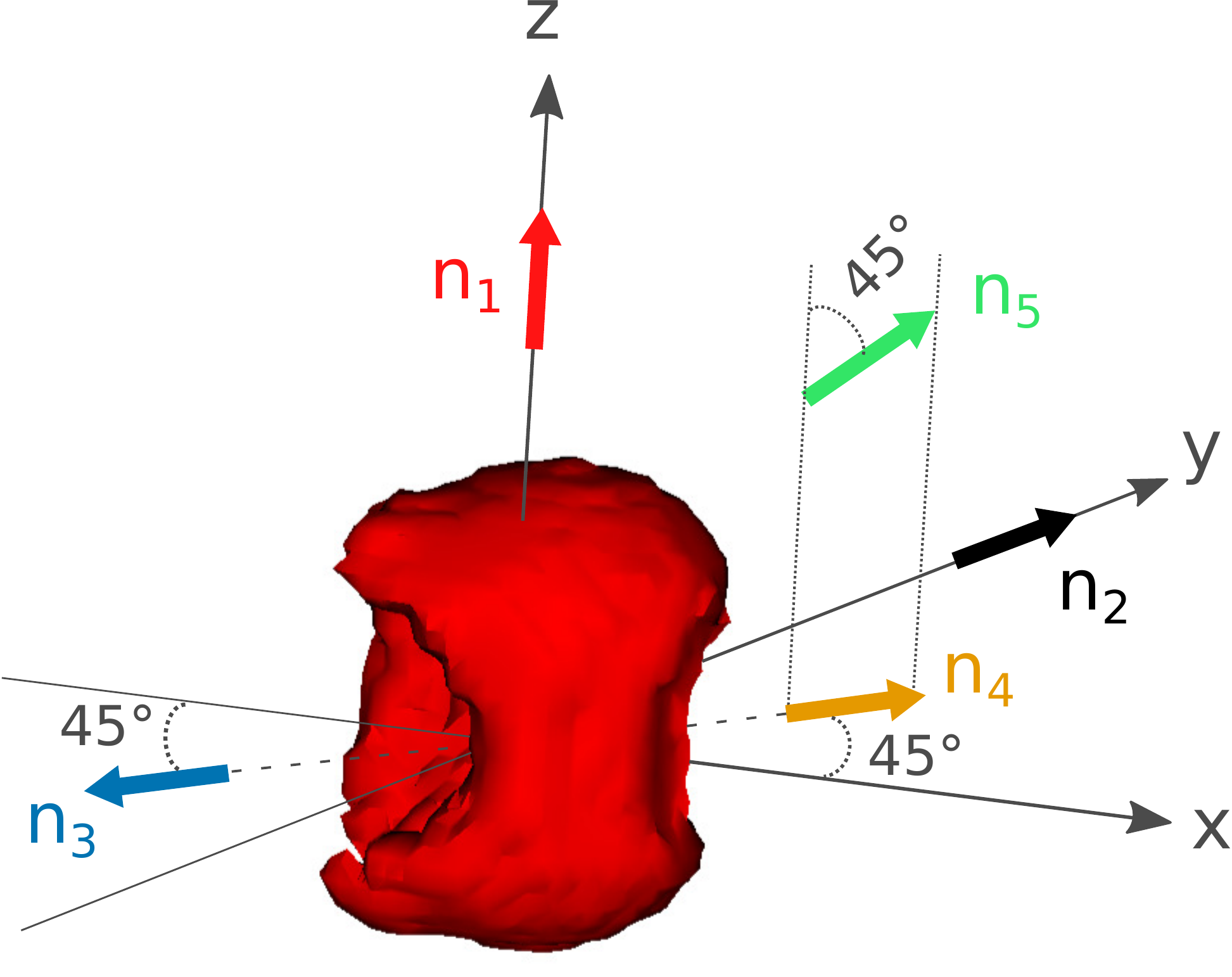}
\caption{Sketch depicting the orientations of the five observers selected in this work with respect to the $^{56}$Ni distribution. Viewing angle $\bmath{n_2}$, $\bmath{n_3}$ and $\bmath{n_4}$ are in the equatorial plane, with $\bmath{n_3}$ facing the cavity created by the presence of the secondary star (see text). }
\label{observers}
\end{center}
\end{figure}

Synthetic observables presented in this paper are extracted exploiting a method recently implemented \mbox{in~\textsc{artis}}~\citep{bulla2015}. In this technique (\textit{event-based technique}, EBT), viewing angle observables are calculated by summing contributions from virtual packets created at each Monte Carlo quanta interaction point and forced to escape to a selected observer orientation. Using the EBT, here we compute intensity and polarisation spectra for virtual packets escaping between 10 and 30 days after explosion with rest frame wavelength in the range 3500$-$10\,000~\AA. Aiming to investigate the impact of aspherical ejecta on the observables, we select five different observer orientations (see Fig. \ref{observers}). Guided by the findings of \citet{pakmor2012}, we first choose the two orientations for which the model is faintest, $\bmath{n_1}=(0,0,1)$, and brightest, \mbox{$\bmath{n_2}=(0,1,0)$}. We then place two observers with orientations that are aligned with the cavity in the IGE distribution that is due to material from the secondary (see Section \ref{model}): one facing the cavity, $\bmath{n_3}=(-\nicefrac{1}{\sqrt{2}},-\nicefrac{1}{\sqrt{2}},0)$, and one at the opposite side, $\bmath{n_4}=(\nicefrac{1}{\sqrt{2}},\nicefrac{1}{\sqrt{2}},0)$. Finally, we select a viewing angle away from the orbital plane and the polar axis, specifically $\bmath{n_5}=(\nicefrac{1}{2},\nicefrac{1}{2},\nicefrac{1}{\sqrt{2}})$. Given that polarisation spectra are typically noisier in the red where the flux is lower, this simulation has been supplemented by an additional calculation with $N=1\times10^8$ and for virtual packets with emergent rest-frame wavelength between 6300 and 10\,000~\AA.

In addition, we also carried out a simulation with \mbox{$N_\text{q}=5\times10^7$} Monte Carlo quanta and with polarisation spectra extracted for an additional 30 viewing angles. Owing to its lower signal-to-noise (a factor of 4 fewer packets than in the previous simulation), the purpose of this calculation is not to study spectral features in details but rather to map out the range of polarisation covered by the model.

\section{Synthetic observables}
\label{synth}

\begin{figure*} 
\begin{center}
\includegraphics[width=1\textwidth,trim=20pt 0pt 0pt 0pt]{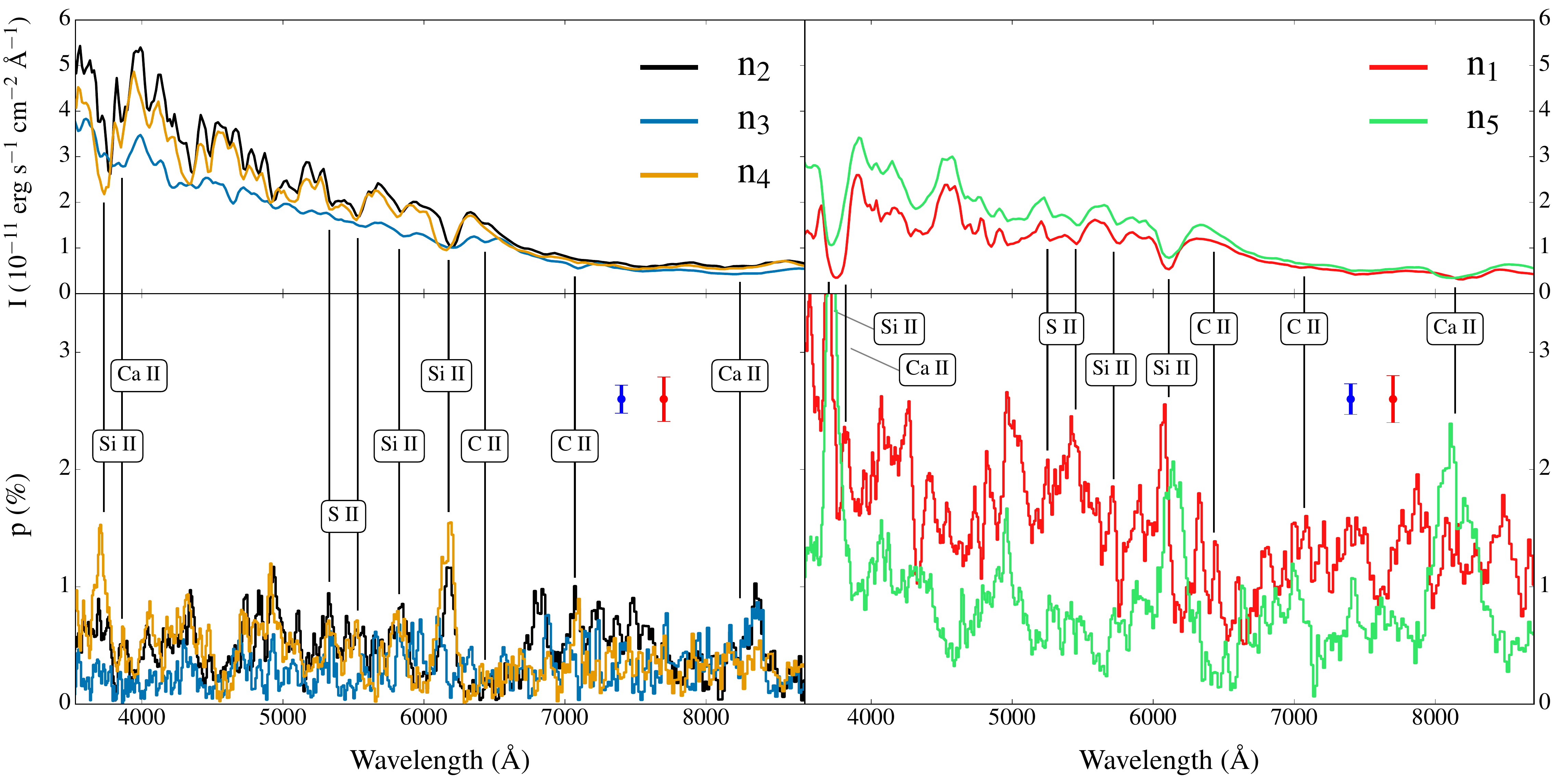}
\caption{Flux and polarisation spectra at $21$ days after explosion for five different viewing angles (see Fig. \ref{observers}). Observers in the orbital plane (black, blue and orange, left-hand panels) typically see brighter and less polarised spectra than those out of the plane (red and light green, right-hand panels). Black vertical lines and labels provide identifications between polarisation peaks and some individual spectral transitions for light and intermediate-mass elements. The error bars mark the averaged Monte Carlo noise level in the polarisation spectra below (blue) and above (red) $6400$ \AA, estimated as in \citet{bulla2015}. Polarisation spectra are Savitzky-Golay filtered on a scale of 3 pixels ($\sim$~40~\AA) for clarity. The model flux is given for a distance of $1$~Mpc.}
\label{specmax}
\end{center}
\end{figure*}

In the following, we present synthetic observables extracted with the EBT for the explosion model introduced in Section~\ref{model}. In Section \ref{polmax} we show flux and polarisation spectra around maximum light for the five observer orientations selected above. We study the spectral evolution in Section \ref{evolution} and finally compare our predictions with spectropolarimetric data of normal SNe Ia in Section \ref{obscompare}. 

\subsection{Polarisation around maximum light}
\label{polmax}

In Fig. \ref{specmax} we report flux and polarisation spectra at 21~days after explosion (around $B$-band maximum light) for the five orientations chosen in Section \ref{radtransf}. As expected, we find the same strong viewing-angle dependences reported for the spectra by \citet[][but see also \citealt{moll2014}]{pakmor2012}. Packets escaping away from the equatorial plane (along $\bmath{n_1}$ or $\bmath{n_5}$) sweep out a larger velocity range in their journey to the observer compared to those escaping in the equatorial plane (see the ejecta distribution in Fig.~\ref{ejecta}) and the corresponding spectra are therefore affected by broader features and stronger blending of lines. In addition, the projected areas of the $^{56}$Ni along $\bmath{n_1}$ and $\bmath{n_5}$ are smaller than those along the chosen equatorial directions and this, together with the higher opacity seen by the escaping packets, translates into fainter spectra. Despite this general trend, the spectrum seen down the cavity (along $\bmath{n_3}$) is fainter than the other two equatorial viewing angles because less IGE material is located on the side near the observer. Compared to the other four orientations, the spectrum extracted along $\bmath{n_3}$ is also markedly different, in the sense that it is characterised by much weaker and less blue-shifted absorption lines. The latter results were also predicted by \citet{kasen2004} for a somewhat similar situation in which a companion star in the single-degenerate scenario carves out a hole in the ejecta of an exploding white dwarf. Unlike our model, however, \citet{kasen2004} adopted a density profile that leads to lines-of-sight through the hole having a lower-than-average column density (see their fig.~1). Therefore, they predict spectra that are typically brighter when viewed through the hole, in contrast with what we see in our model. 

\begin{figure}
\begin{center}
\includegraphics[width=0.473\textwidth,trim=10pt 0pt 0pt 0pt]{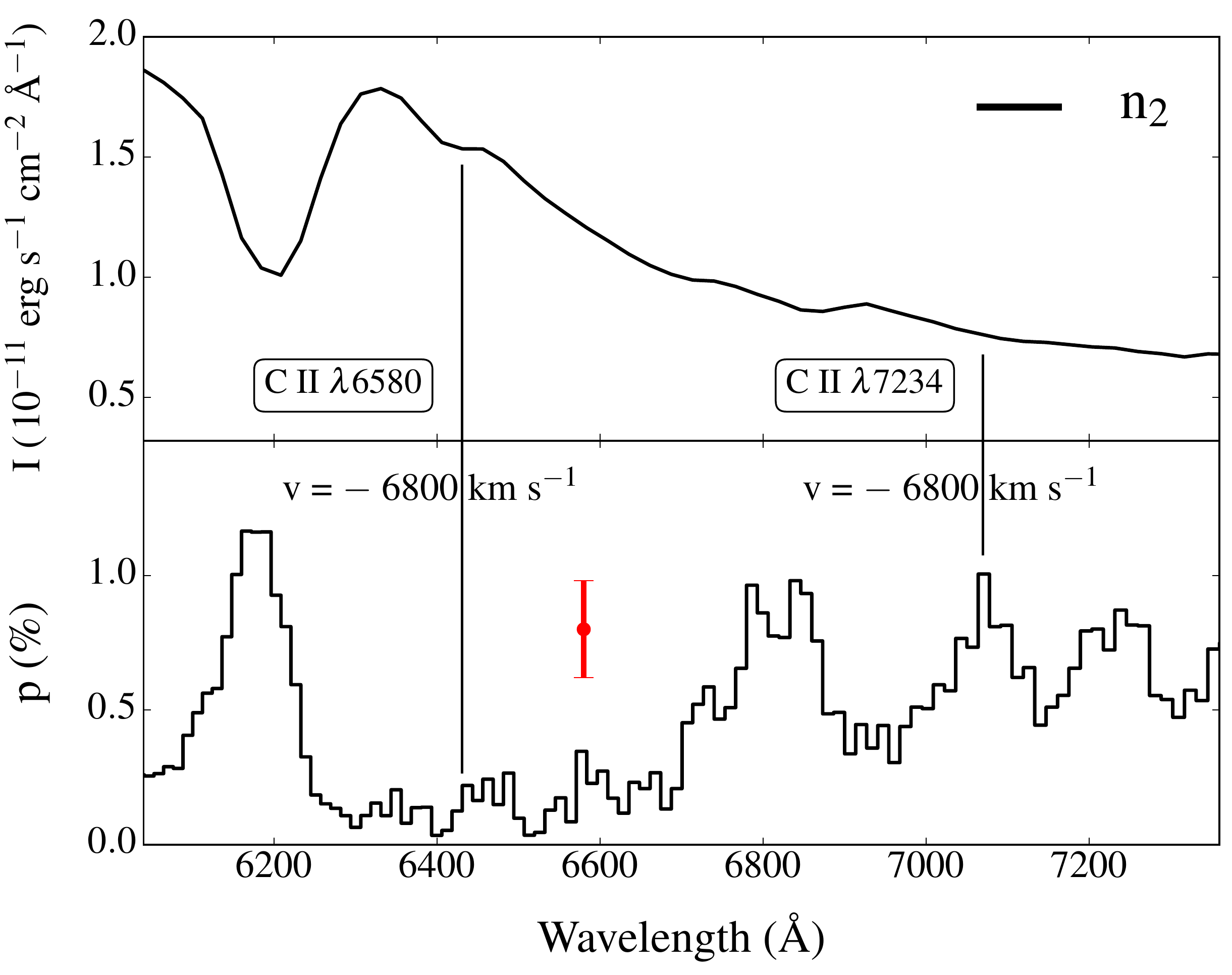}
\caption{Identification of C\,{\sc ii} $\lambda6580$ and C\,{\sc ii} $\lambda7234$. Flux and polarisation spectra are reported at $21$ days after explosion for the viewing angle $\bmath{n_2}$. Vertical lines represent a 6800~km~s$^{-1}$ blue-shift relative to the two carbon features. A red error bar marks the averaged Monte Carlo noise in the spectral range considered. Polarisation spectra are Savitzky-Golay filtered on a scale of 3 pixels ($\sim$~40~\AA) for clarity.}
\label{zoomcarbon}
\end{center}
\end{figure}

As expected from the asymmetric distribution of the ejecta, non-zero polarisation levels are obtained for all the five chosen viewing angles. As shown in Fig.~\ref{specmax}, we find the polarisation signals to be relatively weak in the spectral range between $6400$ and $7200$\,\AA, often studied as a pseudo-continuum region that is devoid of strong line features \citep{patat2009}. Specifically, the pseudo-continuum polarisation level is modest ($\sim$~0.3~per~cent) for orientations in the equatorial plane ($\bmath{n_2}-\bmath{n_4}$), while higher (0.5$-$1~per~cent) for spectra extracted along $\bmath{n_1}$ and $\bmath{n_5}$. In spite of showing weak signals in the pseudo-continuum region, however, polarisation spectra for all the orientations are characterised by strong (1$-$2~per~cent) peaks associated with blue-shifted absorption troughs of spectral lines. While the identification to individual transitions is obvious for some polarisation features (e.g. for Si\,{\sc ii} and Ca\,{\sc ii} lines), it is less clear in the blue region between 4000 and 5000~\AA{} that is affected by stronger line blending.

Compared to the angular binning approach used by \citet{pakmor2012}, the virtual packet technique of \citet{bulla2015} allows us to extract spectra with lower Monte Carlo noise and thus to clearly identify weak lines in the red region of the spectrum. For instance, by combining information from intensity and polarisation spectra we attribute two features in the range 6200$-$7300~\AA{} to C\,{\sc ii}~$\lambda6580$ and C\,{\sc ii}~$\lambda7234$ (see Fig.~\ref{zoomcarbon}). While we see evidence for the C\,{\sc ii}~$\lambda6580$ feature as a notch in the emission wing of the Si\,{\sc ii}~$\lambda6355$ profile, the polarisation counterpart is not visible\footnote{This effect might be due to a balance between polarising contributions from the C~{\sc ii} absorption trough and depolarising contributions from the Si~{\sc ii} emission wing.}. Despite being weak and hard to see in the flux spectrum, however, the C\,{\sc ii}~$\lambda7234$ line does clearly show up in polarisation with $\sim$~1~per~cent level at peak. Both these C\,{\sc ii}~lines are found at rather low velocities ($v\sim$~$-~6800$~km~s$^{-1}$) and are therefore associated with unburned carbon material in inner regions of the ejecta (see Fig.~\ref{ejecta} and discussion in Section~\ref{conclusions}). 
 
\begin{figure}
\begin{center}
\includegraphics[width=0.485\textwidth,trim=20pt 0pt 0pt 0pt]{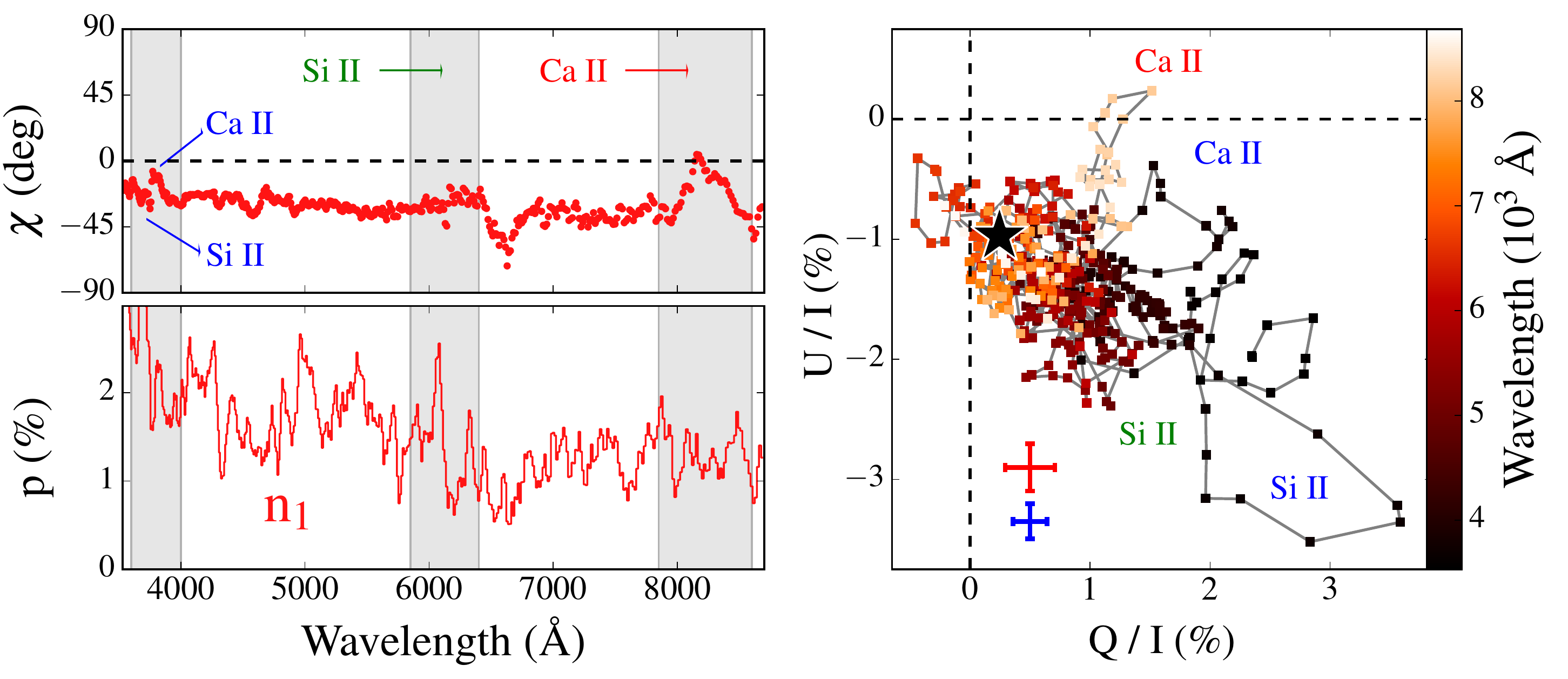}
\includegraphics[width=0.485\textwidth,trim=20pt 0pt 0pt 0pt]{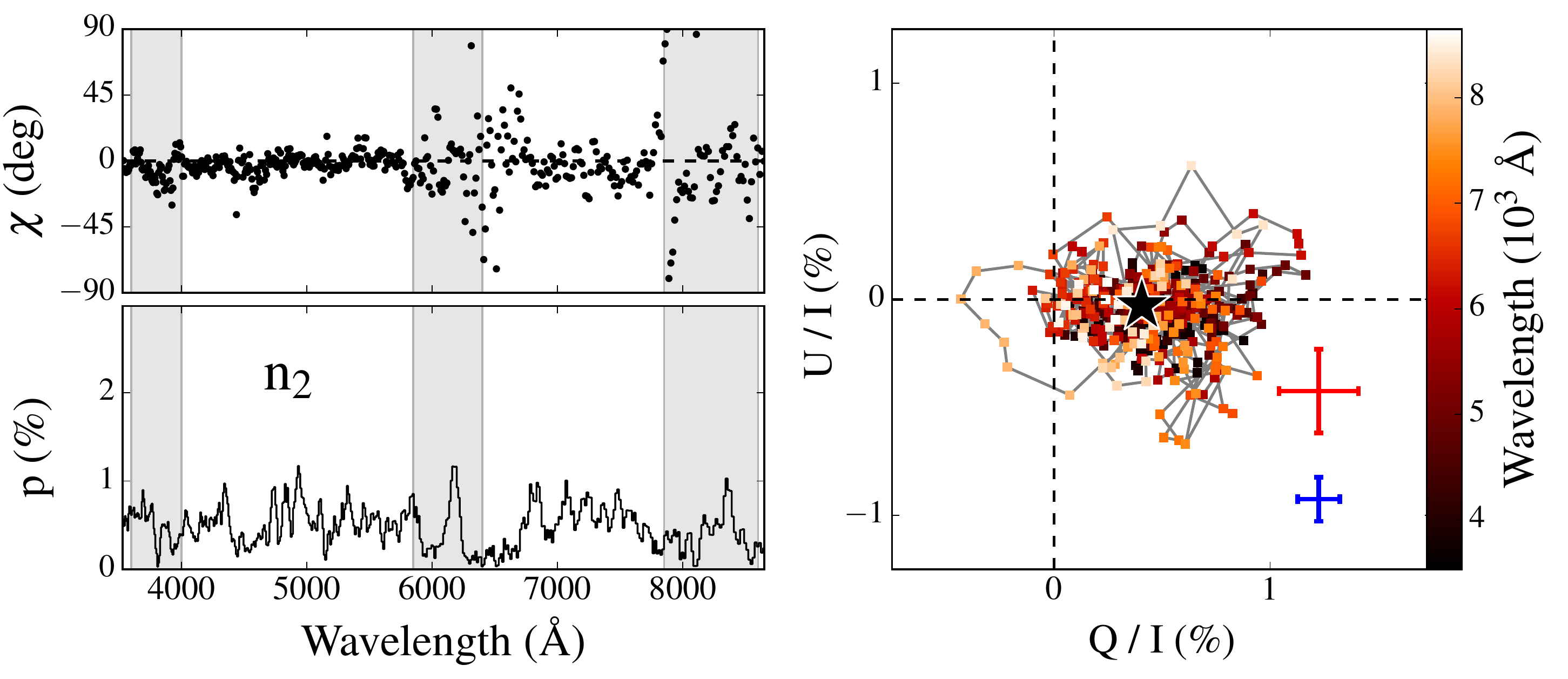}
\includegraphics[width=0.485\textwidth,trim=20pt 0pt 0pt 0pt]{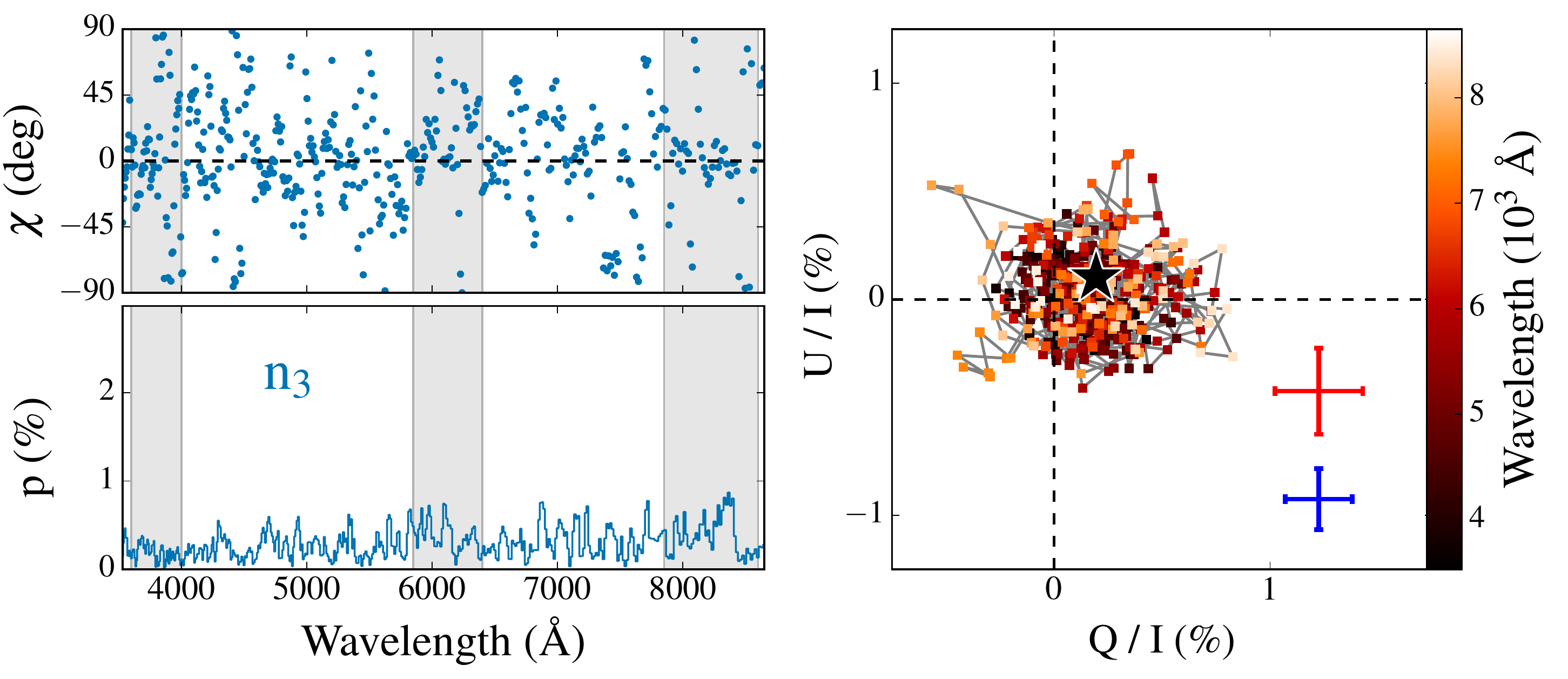}
\includegraphics[width=0.485\textwidth,trim=20pt 0pt 0pt 0pt]{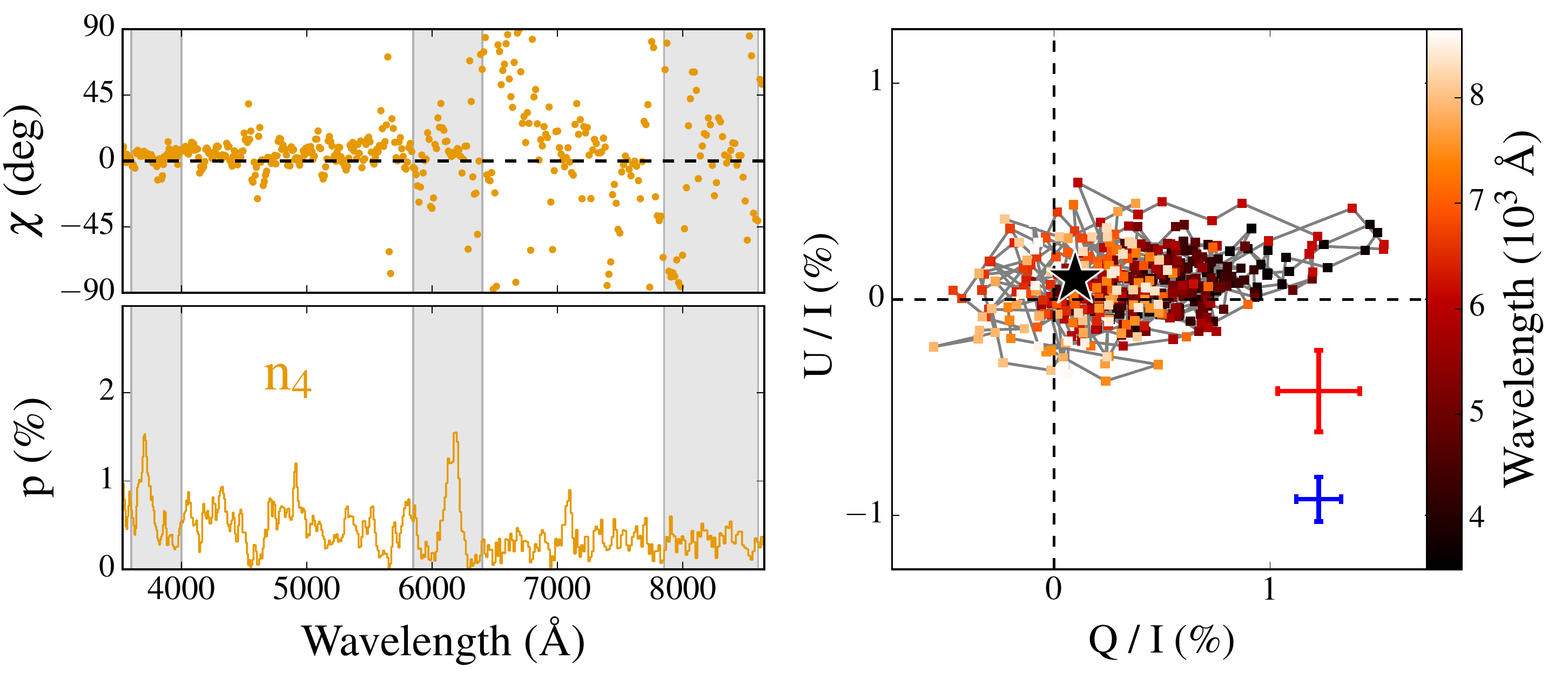}
\includegraphics[width=0.485\textwidth,trim=20pt 0pt 0pt 0pt]{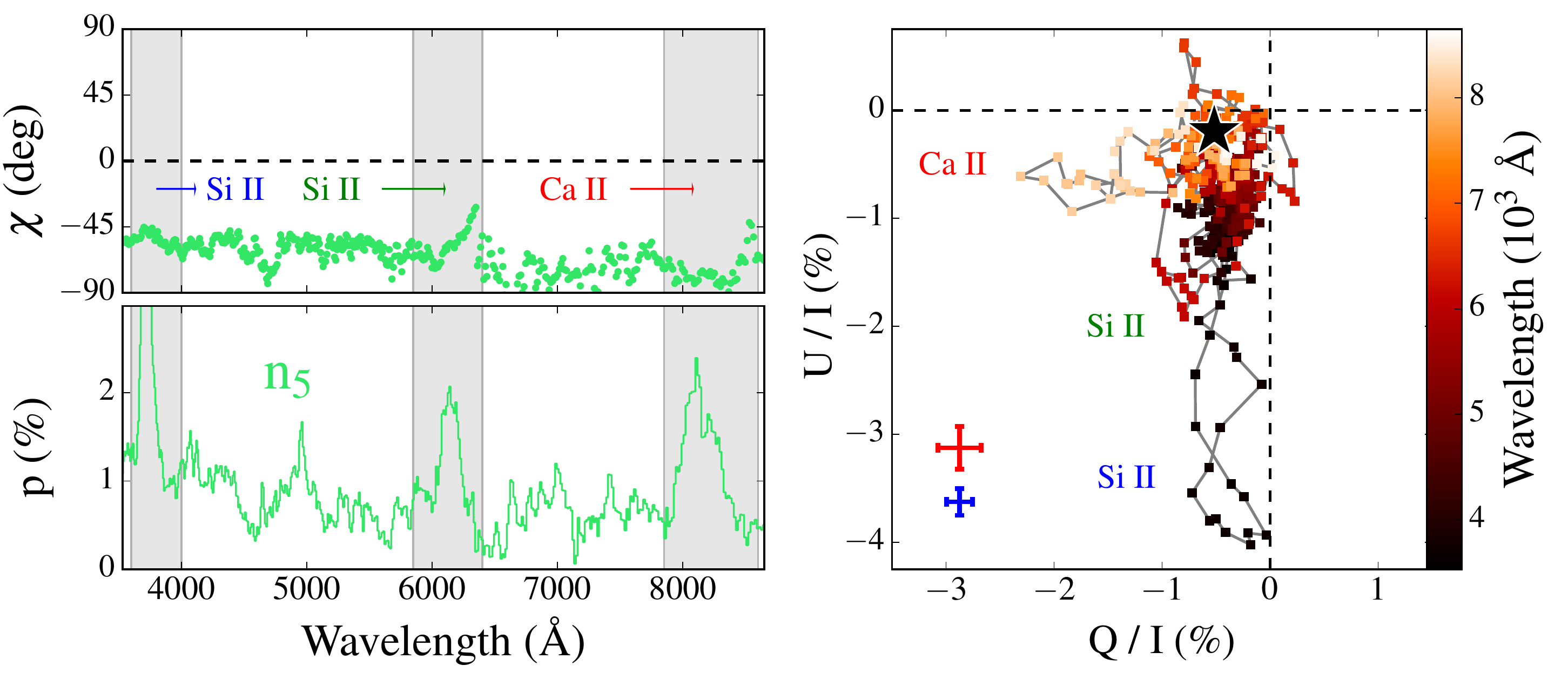}
\caption{Polarisation percentage $p$ and polarisation angle $\chi$ (left-hand panels), together with the corresponding $Q/U$ planes (right-hand panels) for the spectra shown in Fig. \ref{specmax}. For each $Q/U$ plot, a black star indicates the averaged polarisation level in the pseudo-continuum range between $6400$ and $7200$ \AA{}, while the error bars mark the averaged Monte Carlo noise level in the spectra below (blue) and above (red) $6400$ \AA. The Stokes parameters are Savitzky-Golay filtered on a scale of 3 pixels ($\sim$~40~\AA) for clarity. For the $\bmath{n_1}$ and $\bmath{n_5}$ viewing angles we associate individual loops or structures in the $Q/U$ plane to specific features in the spectra.}
\label{quplane}
\end{center}
\end{figure}
 
The $Q/U$ plane provides a valuable tool for linking polarisation spectral features to the ejecta structure: straight lines in this plane suggest the presence of a dominant axis in the ejecta, while deviations from straight lines (e.g. loops) reflect departures from axi-symmetry. This is illustrated in Fig. \ref{quplane}, in which we show the $Q/U$~plane, polarisation level $p$ $\big($=$\sqrt{Q^2+U^2}/I\big)$ and polarisation angle $\chi$ (derived from $\tan2\chi$~=~$U/Q$) for each of the chosen observers. As expected, the symmetry about the orbital plane results in data points distributed along a dominant axis in the $Q/U$ planes for $\bmath{n_2}$, $\bmath{n_3}$ and $\bmath{n_4}$: the up/down symmetry causes every contribution to $U$ from the northern hemisphere to be canceled by a contribution in the southern hemisphere and the $U$ values are therefore consistent with zero. The polarisation angles derived are small which, for these specific orientations, corresponds to an electric field oscillating almost perpendicular to the equatorial plane.
%\footnote{The convention assumes the polarisation angle $\chi$ to run from the north on the sky in the counter-clockwise direction when facing the source.}. 
%Larger polarisation angles are found in the red part of the spectra, but this behaviour reflects the increasing Monte Carlo noise at those wavelengths rather than a departure from the dominant axis. This argument is especially true for the observer facing the hole, $\bmath{n_3}$, for which the signal to noise ratio is poor. 
In contrast, $Q/U$ planes for orientations out of the equatorial plane are characterised by loops associated with different spectral features, mainly Si\,{\sc ii} and Ca\,{\sc ii}. For instance, we see two distinct loops in the blue region below $4000$~\AA{} when the system is viewed pole-on (along $\bmath{n_1}$): the bluer loop is associated with the Si\,{\sc ii} $\lambda3859$ line and shows a polarisation angle comparable to that of the Si\,{\sc ii} $\lambda6355$ feature ($\chi\sim-30^{\circ}$); the redder loop is instead associated with the Ca H and K line and shows a polarisation angle comparable to that of the Ca\,{\sc ii} IR triplet ($\chi\sim-10^{\circ}$).
%Therefore, $Q/U$ planes for orientations out of the plane point toward silicon and calcium morphologies that are distinct (different regions of the $Q/U$ plane, i.e. different polarisation angles) and not perfectly axis-symmetric (loops rather than straight lines).

\begin{figure}
\begin{center}
\includegraphics[width=0.475\textwidth]{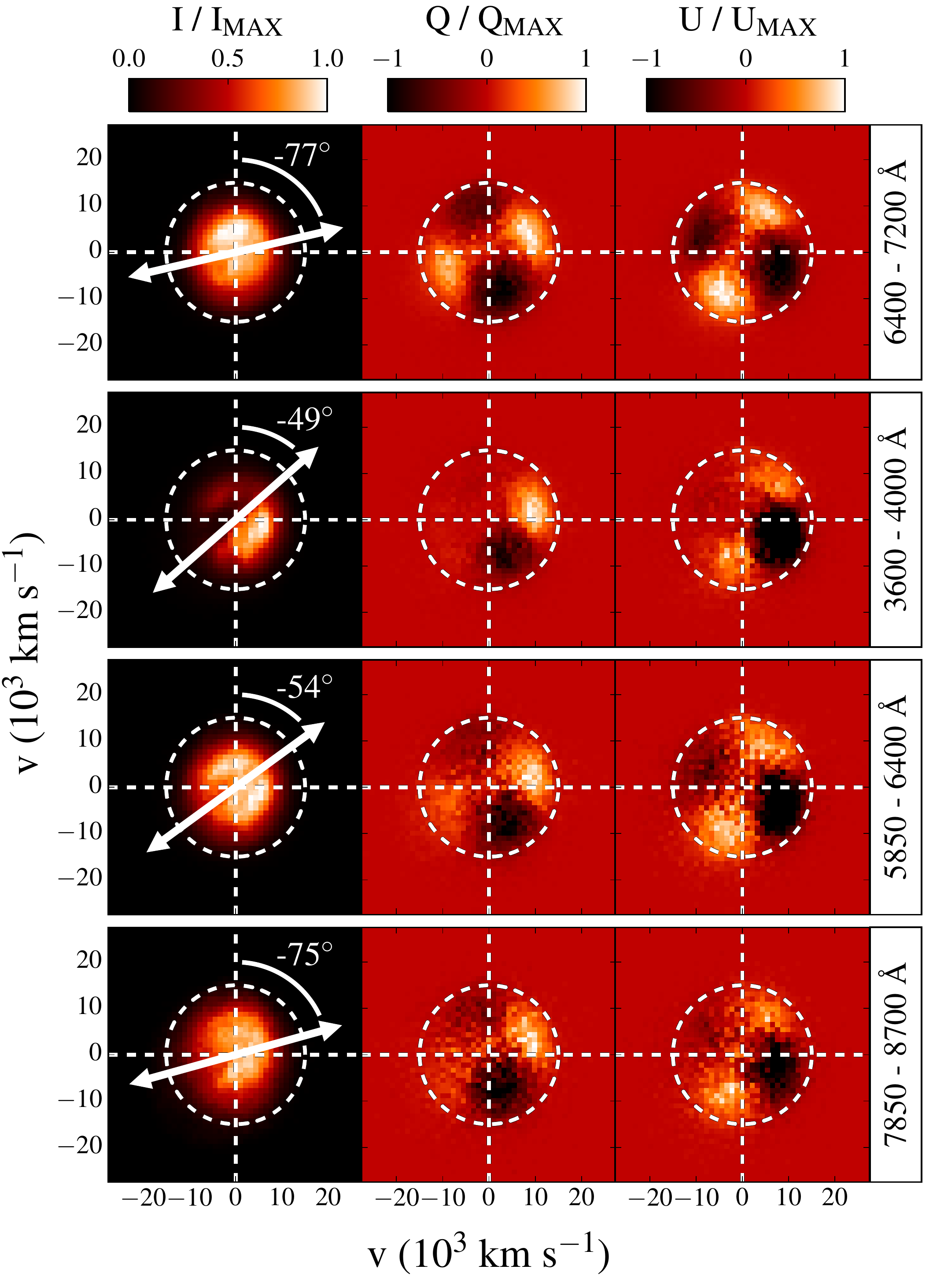}
\caption{Colour maps of normalised $I$ (left-hand panels), $Q$ (middle panels) and $U$ (right-hand panels) distributions projected on the plane perpendicular to the viewing angle $\bmath{n_5}$. The maps are computed selecting virtual packets escaping between $18.5$ and $23.5$ days after explosion and in the wavelength regions 6400$-$7200~\AA{}, 3600$-$4000~\AA, 5850$-$6400~\AA{} and 7850$-$8700~\AA{} (from top to bottom rows). White circles mark a projected velocity of $15~000$~km~s$^{-1}$. White arrows in the $I$ maps indicate the electric field orientation obtained by calculating the polarisation angle $\chi$ from the $Q$ and $U$ maps. The polarisation angles derived for each spectral range are also reported.}
\label{grid_n5}
\end{center}
\end{figure}

\subsubsection{Additional diagnostics}

\begin{figure}
\begin{center}
\includegraphics[width=0.485\textwidth,trim=15pt 0pt 0pt 0pt]{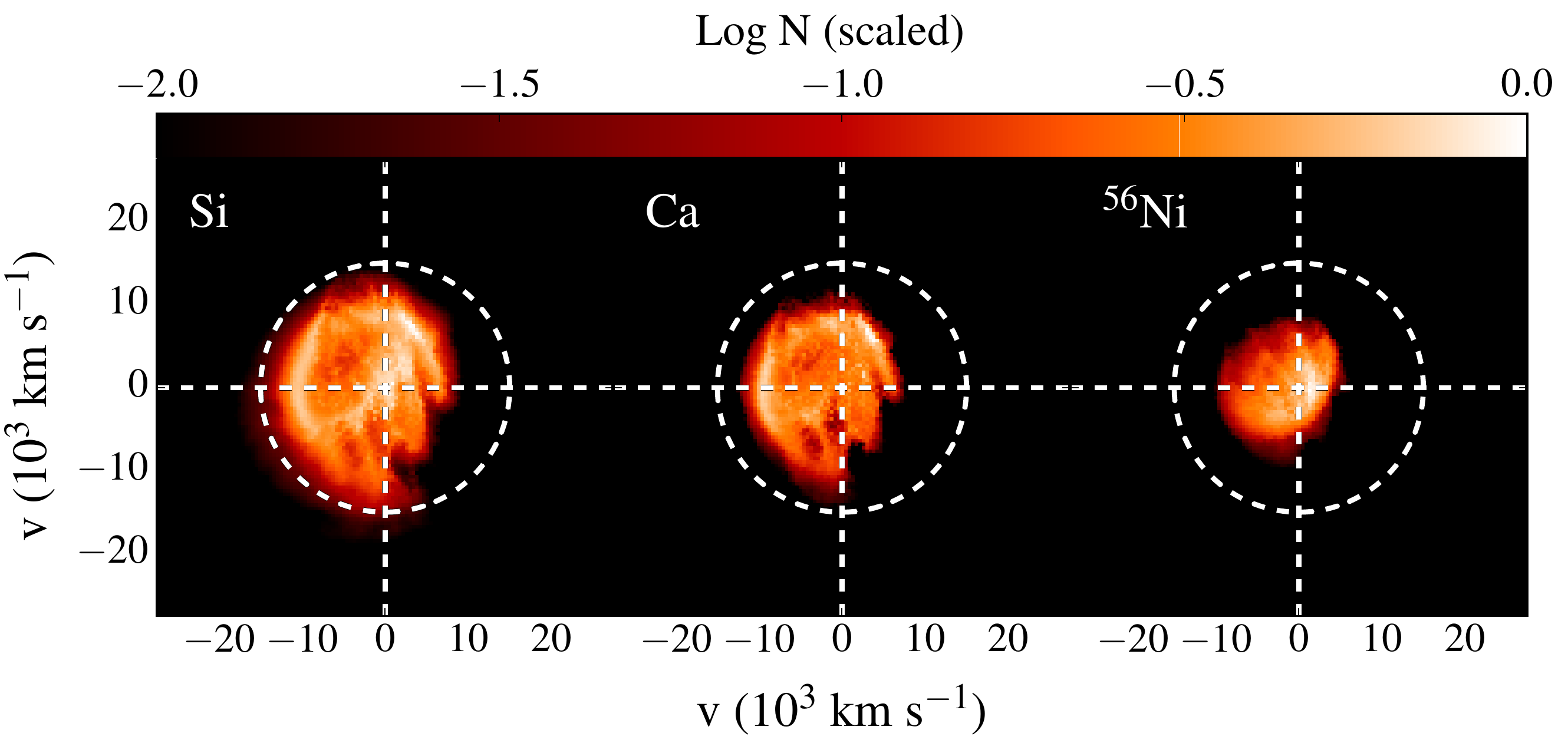}
\caption{Colour maps of the silicon (left-hand panel), calcium (middle panel) and radioactive $^{56}$Ni (right-hand panel) column densities $N$ 100~s after explosion. These are calculated through the near-side hemisphere of the ejecta in the $\bmath{n_5}$ direction. The column density of each map has been scaled by the maximum value. White circles mark a projected velocity of $15~000$~km~s$^{-1}$.}
\label{3dmaps}
\end{center}
\end{figure}

To further investigate how asymmetries in the element distribution are related to individual polarisation spectral features, here we inspect the ejecta morphology of our model more closely. Polarisation levels across spectral lines can be understood as the result of partial covering of the effective electron-scattering photosphere. Fig. \ref{grid_n5} shows the intensity and polarisation distributions projected on the plane perpendicular to the viewing angle $\bmath{n_5}$. The maps have been calculated selecting the emergent virtual packets between $18.5$ and $23.5$ days after explosion and in four different spectral regions: \mbox{6400$-$7200}~\AA{} (pseudo-continuum range), \mbox{3600$-$4000~\AA{}} (comprising the Si\,{\sc ii}~$\lambda3859$ and Ca H and K lines), \mbox{5850$-$6400~\AA{}} (Si\,{\sc ii}~$\lambda6355$) and 7850$-$8700~\AA{} (Ca\,{\sc ii}~IR~triplet). The pseudo-continuum region is almost devoid of lines and therefore more representative of the underlying electron-scattering photosphere. This is relatively symmetric in projection and the corresponding degree of polarisation in this range is close to zero. Silicon and calcium are distributed preferentially on the left side of the ejecta as viewed from this observer orientation (see Fig.~\ref{3dmaps}), and thus the intensity and polarisation maps of Fig. \ref{grid_n5} are brighter for regions on the right where the column densities are lower and from which the photons can leak out more easily.  As a result, the three spectral ranges associated with calcium and silicon lines are polarised in the negative $Q$-direction and negative $U$-direction. Specifically, the polarisation peak in the range 3600$-$4000~\AA{} can be associated with Si\,{\sc ii}~$\lambda3859$ since the polarisation angle inferred from the $Q/U$ map ($\chi=-49^{\circ}$) is consistent with that of Si\,{\sc ii} $\lambda6355$ ($\chi=-54^{\circ}$) and we do not see evidence for a clear Ca H~and~K signature in either the polarisation spectrum (see comparison with the observer $\bmath{n_1}$ in Fig.~\ref{specmax}) or the $Q/U$ plane. However, the Ca\,{\sc ii} IR triplet does show up clearly and has a different polarisation angle ($\chi=-75^{\circ}$), reflecting the different distribution of calcium and silicon in the ejecta.

%Strong polarisation in calcium and silicon but not in oxygen -> as observed. Oxygen more spherically symmetric (see discussion patat 2009 for sn2006x)

\subsection{Spectral evolution}
\label{evolution}

In this section we present the spectral time evolution for the violent merger model of \cite{pakmor2012}. To explore the range of polarisation covered by the model, here we also include results of our simulation with $5\times10^7$ Monte Carlo quanta and with polarisation spectra extracted for an additional 30 viewing angles (see Section \ref{radtransf}). In Section \ref{pol_lc_section} we discuss the polarisation evolution over the time interval between 10 and 30 days after explosion, while in Section~\ref{polvsbrightness_section} we investigate if changes in polarisation for different orientations correlate with light curve properties.

\subsubsection{Polarisation light curves}
\label{pol_lc_section}

\begin{figure}
\begin{center}
\includegraphics[width=0.47\textwidth]{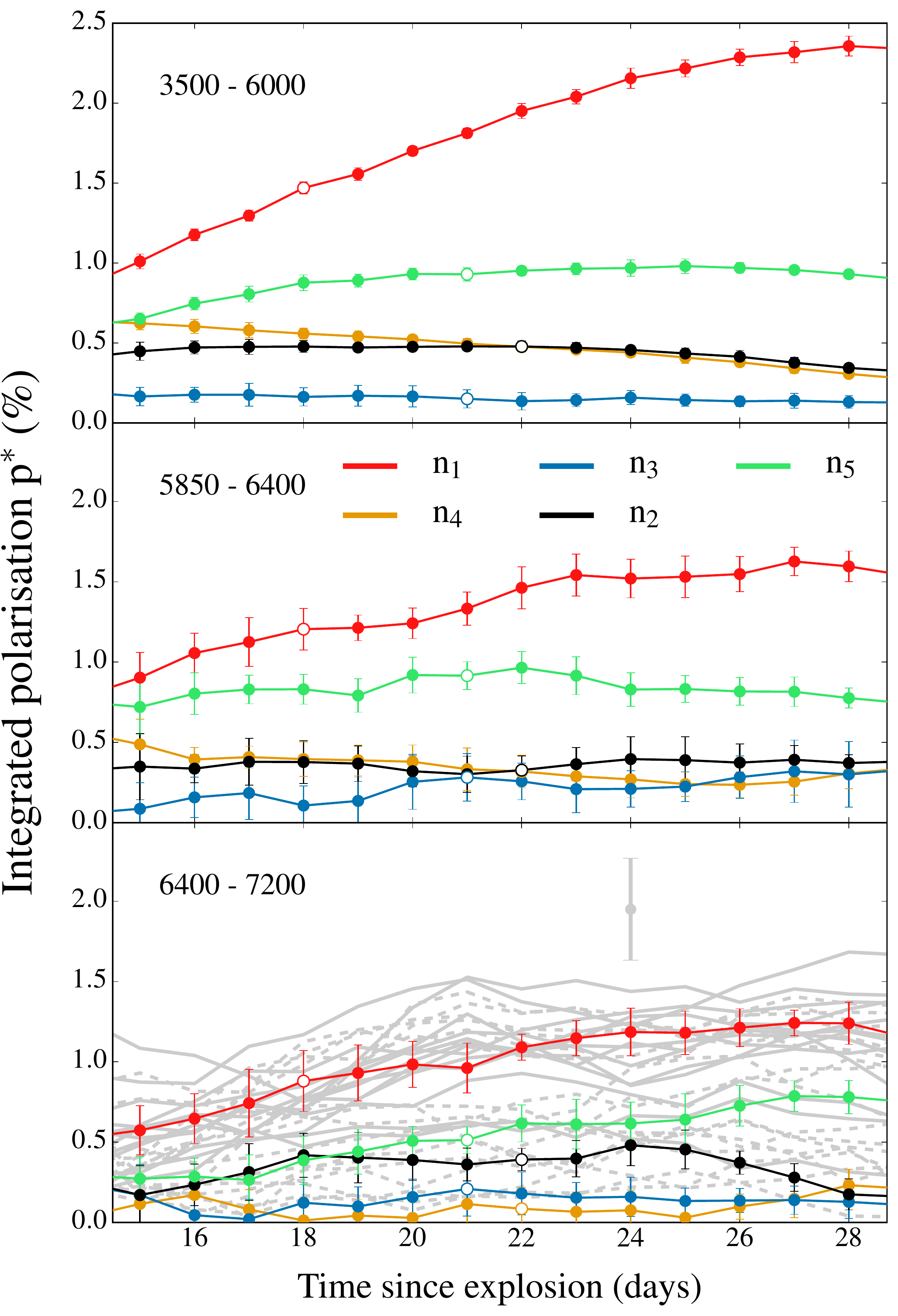}
\caption{Polarisation light curves between 15 and 28 days after explosion calculated in three different spectral range: 3500$-$6000~\AA{} (top panel), 5850$-$6400~\AA{} (middle panel) and 6400$-$7200~\AA{} (bottom panel). Different colours refer to different viewing angles and the colour scheme for the observers $\bmath{n_1}-\bmath{n_5}$ is the same of Fig. \ref{specmax}. An additional 30 light curves (grey lines) are reported in the bottom panel and refer to a lower signal-to-noise (a factor of 4 fewer packets) calculation: solid lines are for observer orientations in a solid angle $d\Omega=2\pi$ around the cavity in the IGE ejecta, while dashed lines stand for orientations on the opposite hemisphere. 1$\sigma$ Monte Carlo noise error bars are derived following \citet{bulla2015} and are larger for simulations carried out with fewer packets  (grey lines). Open points mark the closest epoch to $B$-band maximum for the observers $\bmath{n_1}-\bmath{n_5}$. }
\label{vlcpol}
\end{center}
\end{figure}

\begin{figure}
\begin{center}
\includegraphics[width=0.48\textwidth,trim=7pt 0pt 0pt 0pt]{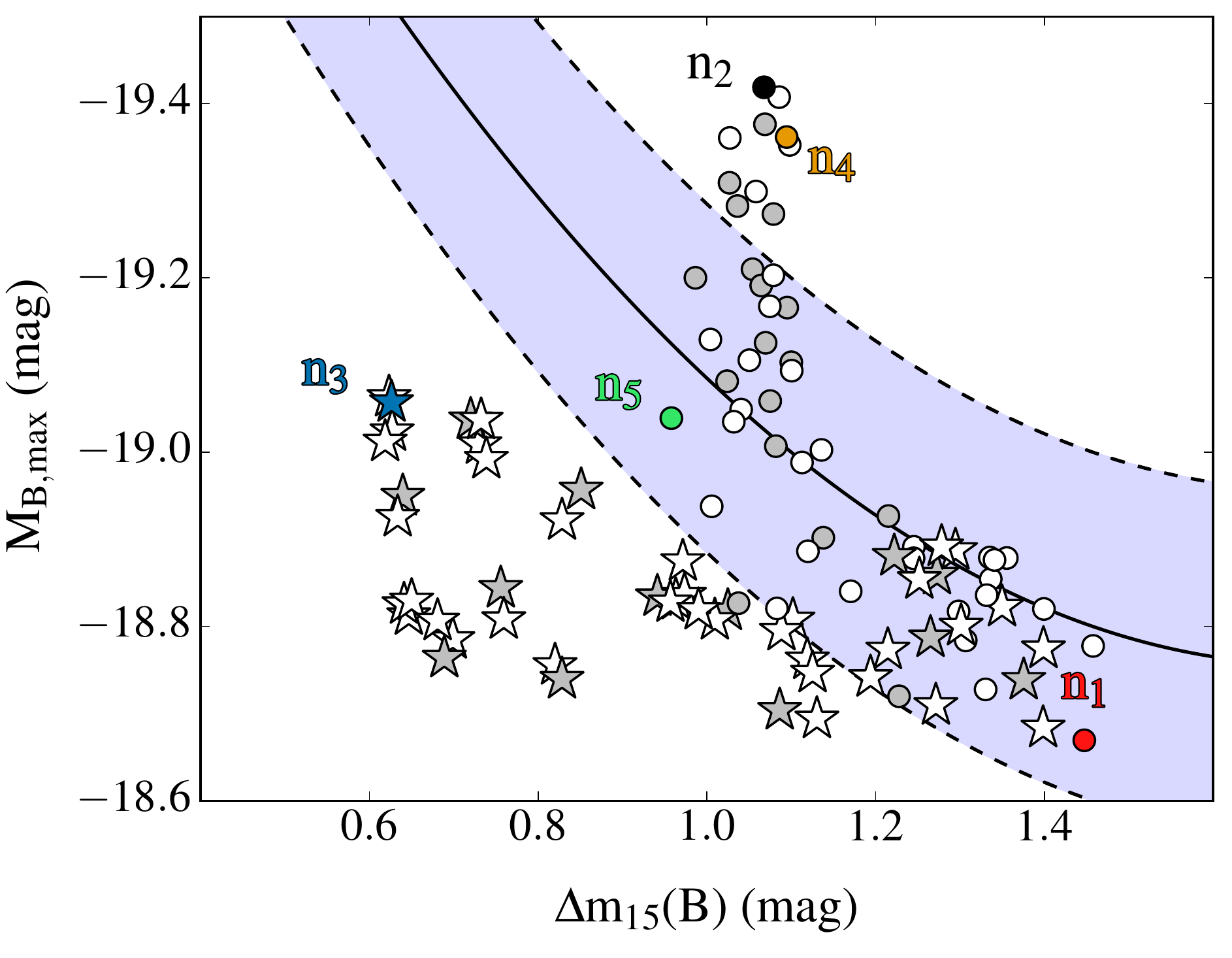}
\caption{Width-luminosity relation in our simulations. Individual points represent different orientations and are calculated by binning the escaping packets into 100 solid angle bins. Angular bins corresponding to the five orientations $\bmath{n_1}-\bmath{n_5}$ are shown with the same colour scheme used in Fig. \ref{specmax}, while grey points mark 30 orientations from a lower signal-to-noise calculation. Open symbols are points for which we did not carry out polarisation synthesis. Points are also divided according to whether they are located in a solid angle $d\Omega=2\pi$ around the IGE cavity (stars) or on the opposite side (circles). The observed trend from \citet{phillips1999}, $M_\textrm{B,max}= M_0 + 0.786\,[ \Delta m_{15}(B)-1.1] - 0.633\,[ \Delta m_{15}(B)-1.1]^2$, is shown with black lines. Different lines refer to different normalisation factors $M_0$ at $\Delta m_{15}(B)=1.1$: $M_0=-19.0$ (solid line), $M_0=-19.2$ (upper dashed line) and \mbox{$M_0=-18.8$} (lower dashed line).}
\label{phillips}
\end{center}
\end{figure}

To investigate how the polarisation signal evolves with time, we calculate intensity ($I^*$) and polarisation ($Q^*$ and $U^*$) light curves by integrating the Stokes parameters over chosen wavelength regions ($\lambda_1$ to $\lambda_2$)
\begin{equation}
\begin{bmatrix} I^*(t) \\ Q^*(t) \\ U^*(t) \end{bmatrix} = \bigintsss_{\lambda_1}^{\lambda_2} \begin{bmatrix} I(\lambda,t) \\ Q(\lambda,t) \\ U(\lambda,t) \end{bmatrix}  ~\mathrm{d}\lambda~,
\end{equation}
from which we derive an integrated polarisation level
\begin{equation}
p^*(t) = \frac{\sqrt{Q^*(t)^2+U^*(t)^2}}{I^*(t)}~.
\end{equation}
In Fig. \ref{vlcpol} we show polarisation light curves for the observer orientations $\bmath{n_1}-\bmath{n_5}$ in three different spectral regions: between $\lambda_1=3500$~\AA{} and $\lambda_2=6000$~\AA{}, between $\lambda_1=5850$~\AA{} and $\lambda_2=6400$~\AA{} (around the Si\,{\sc ii} $\lambda6355$ line) and between $\lambda_1=6400$~\AA{} and $\lambda_2=7200$~\AA{} (in the pseudo-continuum range). The highest polarisation levels in all the spectral regions are found for observers out of the equatorial plane ($\bmath{n_1}$ and $\bmath{n_5}$) at all times. Unlike what is seen for the other orientations, the polarisation level along $\bmath{n_1}$ is very strong and still increasing about 10~days after maximum. Such behaviour, not usually observed in SNe Ia (but see also Section~\ref{polvsbrightness_section} and Section~\ref{obscompare}), is found for almost half of the 35 viewing angles investigated in this work, predominantly for observers in a solid angle of $d\Omega\sim2\pi$ around the IGE cavity (see bottom panel of Fig.~\ref{vlcpol}). Specifically, we find light curves similar to those extracted along $\bmath{n_1}$ for $\sim$~75~per~cent of the orientations on the side of the cavity while only for $\sim$~30~per~cent of the orientations on the opposite side.

\subsubsection{Degree of polarisation and light curve properties}
\label{polvsbrightness_section}

As illustrated by the polarisation light curves (Fig.~\ref{vlcpol}), the highest polarisation signals correspond to observers out of the $x-y$ plane for almost all epochs we considered. The fact that the system is also fainter from these observer orientations suggests that viewing-angle effects might cause a relationship between the degree of polarisation and light curve properties in the model. \citet{wang2007} have suggested that the peak-polarisation of the Si\,{\sc ii}~$\lambda6355$ line inversely correlates with luminosity for normal SNe Ia. Here we then examine if such a correlation could, in part, be driven by orientation effects in the model.

\citet{wang2007} use a light-curve shape parameter, $\Delta m_{15}$(B), as a proxy for the SN luminosity. Specifically, the peak $B-$band absolute magnitude, $M_\textrm{B,max}$, and the magnitude decline in the $B-$band from peak to 15~days after, $\Delta m_{15}$(B), have been shown to well correlate in normal SNe~Ia, in the sense that brighter objects tend to decline more slowly \citep{phillips1993,phillips1999}. Because of the strong asymmetries in the ejecta distribution, the explosion model of \citet{pakmor2012} produces a wide range of $M_\textrm{B,max}$ and $\Delta m_{15}$(B) values. Fig.~\ref{phillips} reports these two parameters for light curves extracted by binning the emergent Monte Carlo quanta in 100 solid angle bins. Only half of the points in this plot follow a trend close to the observed width-luminosity relation. In particular, about 70~per~cent of the observers in a solid angle $d\Omega=2\pi$ around the cavity deviate from the expected trend, suggesting that the influence of the secondary star might be negative on the model. In making comparisons of light curve properties, in the following we will therefore (i) use $M_\textrm{B,max}$ as the proxy for light curve properties (since $\Delta m_{15}$(B) is known to be quite sensitive to the specific ionisation treatment; see e.g. \citealt{kromer2009} and \citealt{dessart2014}) and (ii) distinguish the points on the cavity side by using different symbols.

\begin{figure}
\begin{center}
\includegraphics[width=0.475\textwidth,trim=10pt 0pt 0pt 0pt]{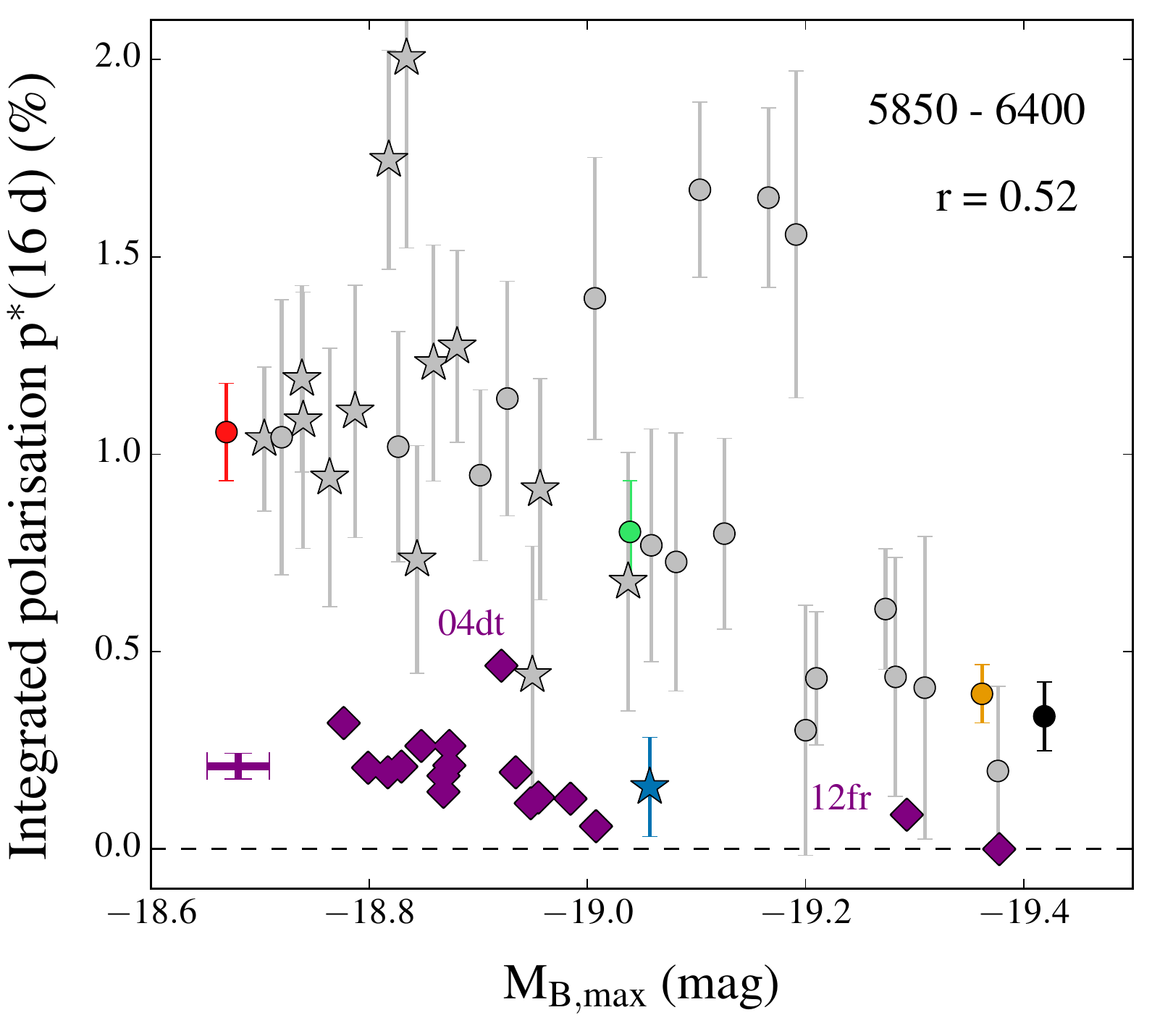}
\caption{Integrated polarisation across the Si\,{\sc ii}~$\lambda6355$ profile as a function of $M_\textrm{B,max}$ (circles and stars). Polarisation levels are extracted in the spectral range 5850$-$6400 \AA{} at 16~days after explosion (about 5 days before $B$-band maximum). Individual points correspond to 35 different viewing angles and are marked and coloured as in Fig.~\ref{phillips}. Error bars are estimated as in \citet{bulla2015} and are larger for simulations that were carried out with a factor of 4 fewer packets (grey points). A Pearson coefficient $r$ calculated including all the points is reported. After being properly converted (see the text), spectropolarimetric data of normal SNe Ia \citep{wang2007,maund2013} are also plotted for comparison (purple diamonds). Purple error bars mark the averaged uncertainties for these values.}
\label{polvsbrightness}
\end{center}
\end{figure}

To look for trends with $M_\textrm{B,max}$, in Fig. \ref{polvsbrightness} we examine polarisation levels integrated across the Si\,{\sc ii}~$\lambda6355$ line (between $5850$ and $6400$ \AA) at 16~days after explosion, $p^*$(16~days). This epoch is chosen as we aim to compare our predictions with the trend observed for normal SNe~Ia 5~days before $B-$band maximum \citep{wang2007}. To facilitate this comparison, we also convert the quantities presented by \citet{wang2007} in the following way: (i) we translate $\Delta m_{15}$(B) into $M_\textrm{B,max}$ values using the \textit{observed} width-luminosity relation reported in Fig.~\ref{phillips} \citep[$M_0=-19.0$;][]{phillips1999}; (ii) we transform the peak-polarisation of the Si\,{\sc ii}~$\lambda6355$ line into an integrated level between $5850$ and $6400$~\AA{} assuming a Gaussian profile with full width at half maximum of 150~\AA\footnote{This full width at half maximum is chosen since it is a good approximation for the values predicted by our model (which span a range between 100 and 200~\AA, see Fig.~\ref{specmax}) and observed in SNe~Ia (see e.g. Fig.~\ref{2004dt}).}. As shown by Fig.~\ref{polvsbrightness}, the correlation between luminosity and degree of polarisation is moderate in our model (Pearson coefficient r~=~0.52) but is in the same sense as is suggested by the data: brighter viewing angles are typically less polarised, fainter viewing angles typically more polarised. The degrees of polarisation predicted by our model, however, are typically stronger than those observed and this discrepancy is found to be more pronounced for orientations on the side of the cavity.

\subsection{Comparison with SN~2004dt and SN~2012fr}
\label{obscompare}

Focusing on high signal-to-noise calculations, in this Section we compare spectra extracted along $\bmath{n_2}-\bmath{n_5}$ with spectropolarimetric data of SNe~Ia. As shown in Section~\ref{polmax} and Section~\ref{evolution}, the degree of polarisation for the model of \citet{pakmor2012} is low ($\lesssim$~1~per~cent) when the system is viewed equator-on ($\bmath{n_2}$, $\bmath{n_3}$ and $\bmath{n_4}$) and higher for the viewing angle out of the equatorial plane ($\bmath{n_5}$). Therefore, in the following we present a comparison of our synthetic spectra with two normal SNe Ia showing different degrees of polarisation: the highly polarised SN~2004dt (which we compare with predictions for the line-of-sight out of the equatorial plane) and the lowly polarised SN~2012fr (which we compare to our calculations in the equatorial plane).

\subsubsection{SN~2004dt}

\begin{figure*}
\begin{center}
\includegraphics[width=1\textwidth]{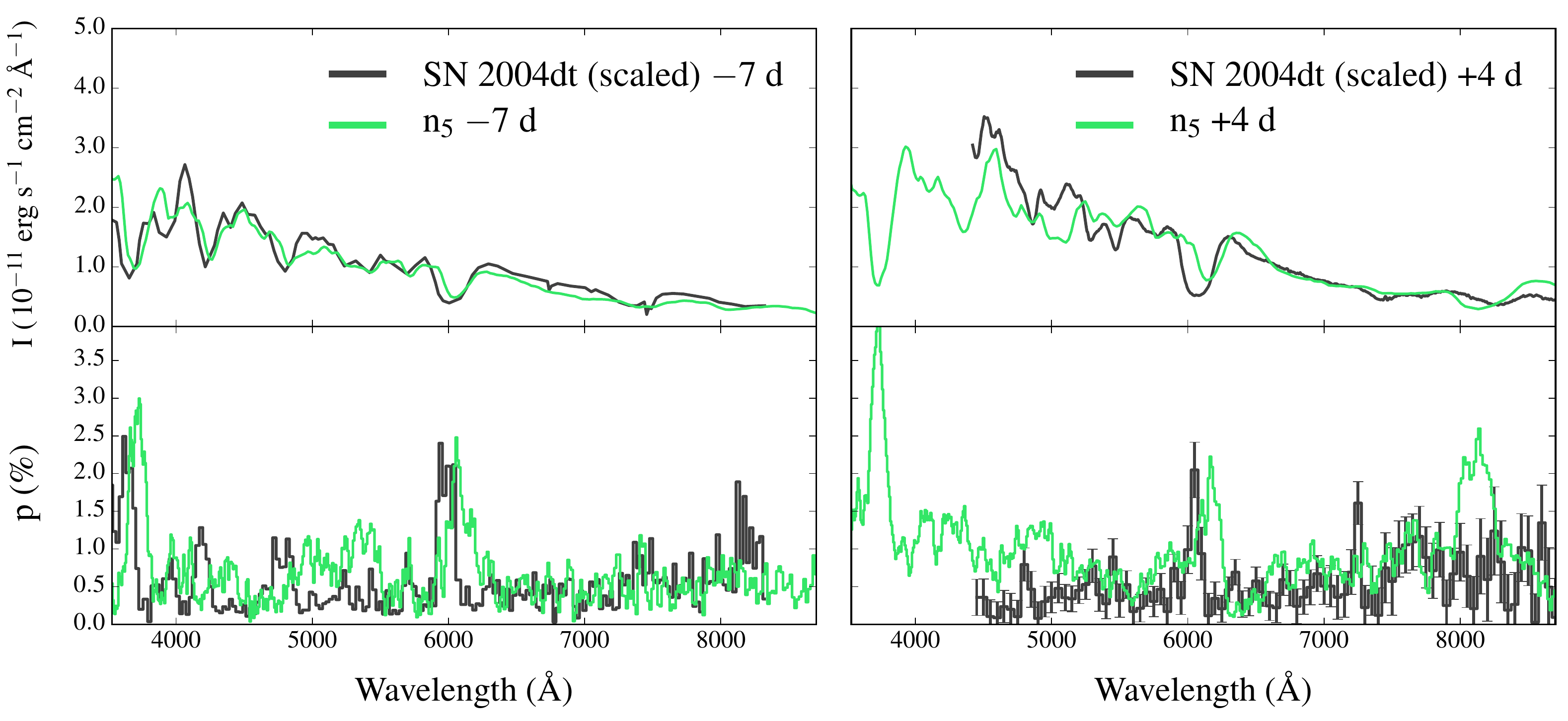}
\caption{Flux and polarisation spectra predicted for an observer orientation $\bmath{n_5}=(\nicefrac{1}{2},\nicefrac{1}{2},\nicefrac{1}{\sqrt{2}})$. Spectra at $-$7 (left panels) and +4 days (right panels) relative to $B$-band maximum are shown with green lines. For comparison, black lines show observed spectra of SN~2004dt at the same epochs after correction for the interstellar polarisation component \citep[$\sim$~0.28~per~cent;][]{wang2006,leonard2005}. Synthetic flux spectra are given for a distance of $1$~Mpc, while flux spectra of SN~2004dt have been arbitrarily scaled for illustrative purposes.}
\label{2004dt}
\end{center}
\end{figure*}

Spectropolarimetric observations of SN~2004dt were triggered with FORS1 at ESO VLT 6$-$8~days before maximum \citep{wang2006} and with the Shane telescope at Lick Observatory 4~days after maximum \citep{leonard2005}. The polarisation signal detected across the Si\,{\sc ii}~$\lambda6355$ line was very prominent ($2-2.5$~per~cent) and little evolution was observed between the two epochs. 

In Fig.~\ref{2004dt} we show observations of SN 2004dt together with synthetic spectra extracted at the corresponding epochs for the viewing angle $\bmath{n_5}$. Although the velocities in the model along $\bmath{n_5}$ are lower than those observed (e.g. $\sim-14~500$~km~s$^{-1}$ instead of $\sim-17~000$~km~s$^{-1}$ for the absorption trough of the Si\,{\sc ii}~$\lambda6355$ line 7~days before maximum), the simulated spectral shapes agree reasonably well with the data of SN~2004dt at both epochs. In particular, the agreement a week before maximum is remarkable. The polarisation level in the pseudo-continuum range 6400$-$7200~\AA{} is well reproduced and the predicted line polarisation features are very similar to those observed. The match is particularly good for the silicon lines (e.g. Si\,{\sc ii}~$\lambda3859$\footnote{Note that the \citet{wang2006} identification of the feature at $3650$ \AA{} with Si\,{\sc ii}~$\lambda3859$ is consistent with our findings of Section~\ref{polmax}.}, Si\,{\sc ii} $\lambda5979$ and Si\,{\sc ii} $\lambda6355$).
% but also reasonable for the Ca\,{\sc ii} IR triplet feature. 
In contrast, the model produces a degree of polarisation that is too high in the region around 5000 \AA{}. 

Synthetic spectra 4~days after maximum are still consistent with data for SN 2004dt, though with stronger deviations. The most striking difference concerns the region across the Ca\,{\sc ii} IR triplet feature, which appears to be too strong in the model. The fact that the calcium line is also prominent in the flux spectrum, however, suggests that the discrepancy might be ascribed to the simplified excitation/ionisation approximation of \textsc{artis} \citep{kromer2009}, rather than only to the geometry of the model. 

\subsubsection{SN~2012fr}

Spectropolarimetric data of SN~2012fr were acquired with FORS1 at ESO VLT at four different epochs: $-$11, $-$5, +2 and +24~days relative to $B$-band maximum \citep{maund2013}. The degree of polarisation of SN~2012fr was generally small, except for the signals associated with high- (early epochs) and low- (late epochs) velocity components of Si\,{\sc ii} $\lambda6355$ and Ca\,{\sc ii} IR lines. \citet{maund2013} argued that the very low level of continuum polarisation seen in SN 2012fr ($<0.1$~per~cent) is not consistent with the asymmetric nickel distribution in the violent white dwarf mergers of \citet{pakmor2012}. Having undertaken a polarisation spectral synthetis for this model, we can now investigate this comparison quantitatively.

\begin{figure*}
\begin{center}
\includegraphics[width=1\textwidth]{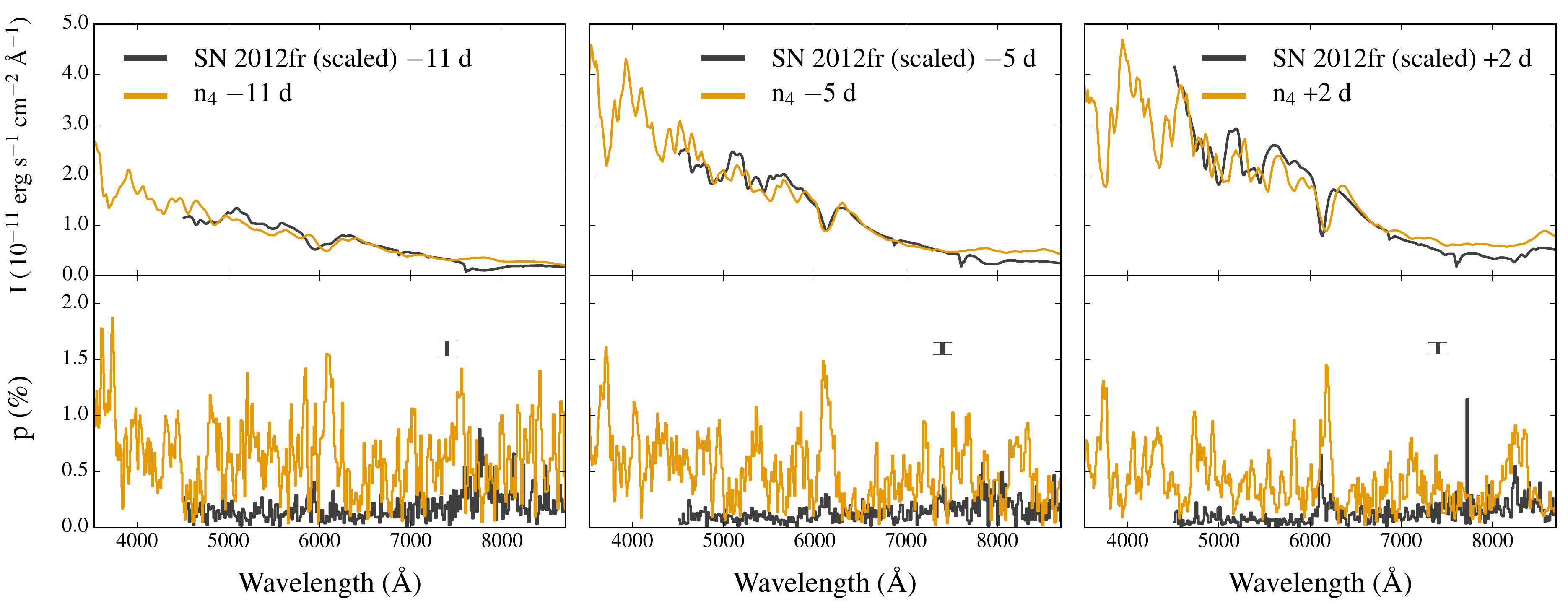}
\caption{Flux and polarisation spectra predicted for an observer orientation $\bmath{n_4}=(\nicefrac{1}{\sqrt{2}},\nicefrac{1}{\sqrt{2}},0)$. Spectra at $-$11 (left-hand panels), $-$5 (middle panels) and +2 days (right-hand panels) relative to $B$-band maximum are shown with orange lines. For comparison, black lines show observed spectra of SN~2012fr at the same epochs after correction for the interstellar polarisation component \citep[$\sim$~0.24~per~cent;][]{maund2013}. Black bars mark the averaged errors in the observed spectra at each epoch. Synthetic flux spectra are given for a distance of $1$~Mpc, while flux spectra of SN~2012fr have been arbitrarily scaled for illustrative purposes.}
\label{2012fr_n4}
\end{center}
\end{figure*}

In Fig.~\ref{2012fr_n4} we show spectropolarimetric data of SN~2012fr together with synthetic spectra extracted in the $\bmath{n_4}$ direction\footnote{We have chosen to compare data with the $\bmath{n_4}$ case here but note that similar conclusions are drawn if we instead use $\bmath{n_2}$, for which the model predicts similar polarisation spectra (see Fig.~\ref{specmax})}. The pseudo-continuum polarisation at all epochs is fairly consistent\footnote{Although the model predicts that several polarised line features do form in the pseudo-continuum region.} with the level observed in SN~2012fr. This suggests that the strong asymmetries in the nickel distribution of \citet{pakmor2012} might not necessarily lead to high polarisation levels in the pseudo-continuum range (see also discussion in Section~\ref{polmax}). However, the most striking discrepancy between the model and the data is the wealth of strong polarisation features predicted in our simulation. As highlighted in Fig.~\ref{specmax} and Fig.~\ref{zoomcarbon}, these strong features are associated with the absorption troughs of spectral lines and are typically too strong to match the modest polarisation signals of SN~2012fr. This fact suggests that the element distribution in the model ejecta is too asymmetric to account for the overall low levels of polarisation observed in SN~2012fr and similar SNe Ia.

\begin{figure*}
\begin{center}
\includegraphics[width=1\textwidth]{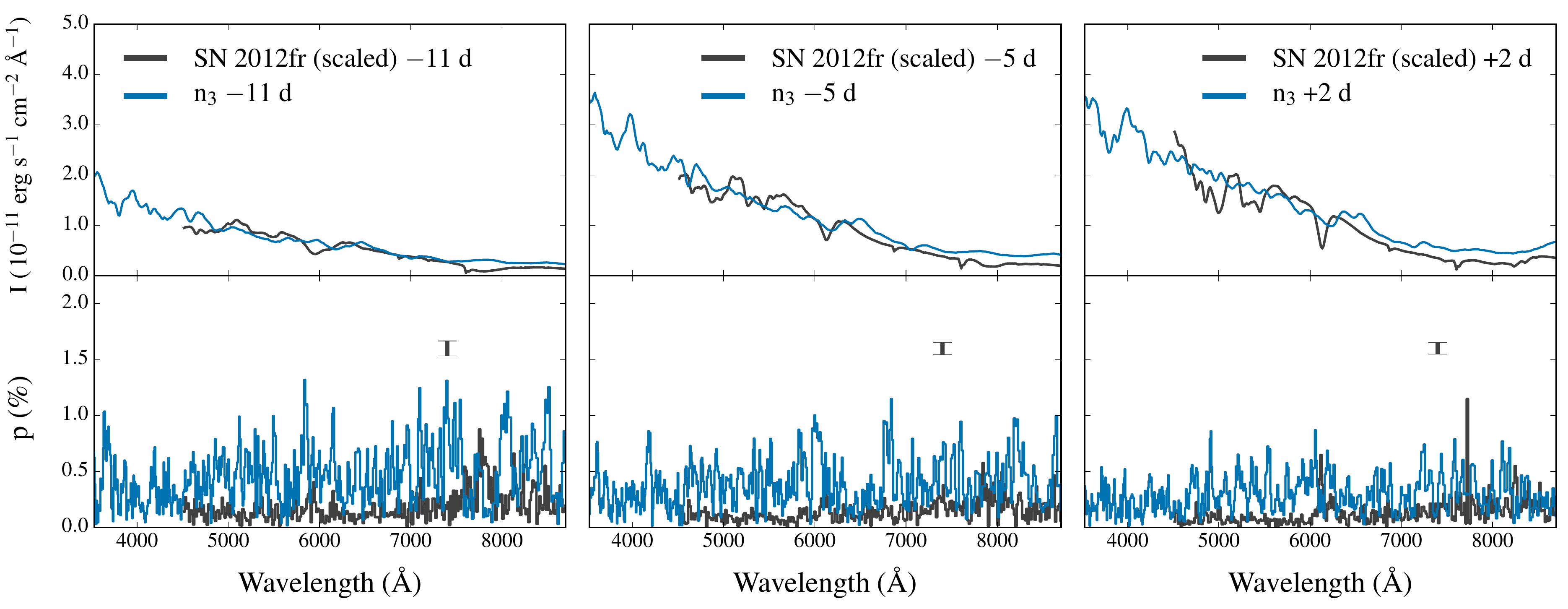}
\caption{Same as Fig~\ref{2012fr_n4} for an observer orientation $\bmath{n_3}=(-\nicefrac{1}{\sqrt{2}},-\nicefrac{1}{\sqrt{2}},0)$.}
\label{2012fr_n3}
\end{center}
\end{figure*}

In Fig.~\ref{2012fr_n3} we compare observations of SN~2012fr with the spectra predicted by our model in the $\bmath{n_3}$ direction. As previously noticed, the degree of polarisation associated with this orientation is typically weaker compared to the other two equatorial orientations that we have considered, and thus it provides better agreement, although still imperfect, if one considers only polarisation. However, the flux spectra extracted for this direction are characterised by very broad and shallow features that are not consistent with the observations of SN~2012fr.

%2004dt. Synthetic spectra do not increase in the red
% See also abstract of Clocchiati+2011: SILICON ("The evolution in time of the feature confirms the picture based on the first spectrum, that the asymmetric silicon structure extends from very high to normal velocities ( 22,000 to 8,000 km/s, although the maximum polarization stays at moderately high velocity ( 13,500 km/s)") -> check if this is true for one of our viewing angle spectra. CALCIUM: ("The line polarization of the low velocity Ca II triplet (at 16,000 km/s), on the other hand, stays at 1.0%, during the photospheric phase and then increases by a factor of more than two, being strongest in our latest observation.") -> It may be that the strong line we see in synthetic spectra (>1%) is because of the ionisation treatment. At that epoch should be still ionised and then it should pop up later on.

% Calcium abundance too high. Pakmor2012: "this offset (in the I band after maximum) can be attributed to a flux excess in the Ca II NIR triplet in the synthetic spectra. This could point to an overabundance of calcium in the model. However, it is more likely to be shortcoming of our radiative transfer treatment, which uses the simple van Regemorter approximation to treat collisional excitation (...)"

\section{Discussion \& conclusions}
\label{conclusions}

We have presented a polarisation spectral synthesis analysis of the violent merger of a 1.1~M$_{\odot}$ and a 0.9~M$_{\odot}$ carbon-oxygen white dwarf \citep{pakmor2012}. Using a technique recently implemented in the Monte Carlo radiative transfer code \textsc{artis} \citep{bulla2015}, we have calculated polarisation spectra for multiple observer orientations between 10 and 30~days after explosion. Our simulations focused on five orientations, for which a large number of Monte Carlo quanta were used in order to extract detailed synthetic observables. Results for those five orientations are supplemented by 30 additional, lower signal-to-noise calculations that allowed us to map out the range of polarisation covered by the model.

Around maximum light, the overall degree of polarisation for viewing angles in the equatorial plane is relatively low ($p\lesssim1$~per cent). Polarisation spectra clearly identify the dominant axis of the model, reflecting the overall symmetry of the ejecta about the $x-y$ plane. In contrast, higher polarisation levels and departures from a single-axis geometry are found for viewing angles out of the equatorial plane. For a given observer, polarisation angles estimated across Si\,{\sc ii} lines differ from those across Ca\,{\sc ii} lines, reflecting the distinct morphologies of silicon and calcium in the ejecta.

Focusing on high signal-to-noise calculations, we compared our synthetic spectra with spectropolarimetric data of two normal SNe Ia: the highly polarised SN~2004dt \citep{leonard2005,wang2006} and the lowly polarised SN~2012fr \citep{maund2013}. Synthetic spectra extracted along the direction $\bmath{n_5}=(\nicefrac{1}{2},\nicefrac{1}{2},\nicefrac{1}{\sqrt{2}})$ provide a good match to those of SN~2004dt at $-$7 and +4 days relative to $B$-band maximum. In particular, the match is remarkable one week before maximum, with the signals across the Si\,{\sc ii} and Ca\,{\sc ii} profiles and in the continuum range found to be very similar to those observed. In contrast, none of the equatorial viewing angles can simultaneously reproduce the polarisation and intensity spectra observed in SN~2012fr. Owing to the strong asymmetries in the element distribution, the model predicts a wealth of strong ($\sim$~1~per~cent) polarisation lines both in the blue and in the red part of the spectrum, in contrast with the behaviour commonly observed in SNe Ia.

In addition, we found that about half of the 35 orientations investigated in this study clearly deviates from the trend observed in SNe~Ia, with polarisation levels that are typically too high and still increasing $\sim$~10 days after maximum light. Interestingly, most of these orientations are oriented close to the cavity in the $^{56}$Ni distribution, which is filled by material from the companion star. The negative impact of the material from the explosion of the companion star on the model is also suggested by the identification in our spectra of two optical C\,{\sc ii} lines at rather low velocities ($v\sim$~$-~6800$~km~s$^{-1}$). These features, associated with unburned carbon material from the companion, are not normally observed at such low velocities \citep[$v_{\text{obs}}\sim$~$-~12\,000$~km~s$^{-1}$;][]{parrent2011,thomas2011} or detected in polarisation \citep[see e.g.][]{wang2008}. 

% or flux spectra of SNe~Ia.

%produces signatures of  at rather low velocity ($\sim$~6800~km~s$^{-1}$). These two C\,{\sc ii} features have been identified in some flux

%for the C\,{\sc ii}~$\lambda6580$ and C\,{\sc ii}~$\lambda7234$ features both in the flux and polarisation spectra

%attributed two C\,{\sc ii}~lines in the range 

%unburned carbon material associated with the companion star produces 
 
With this paper, we have started our project aimed to predict polarisation signatures for a set of contemporary SN~Ia explosion models. We have investigated the particular model of \citet{pakmor2012} and shown that this violent merger of a 1.1~M$_{\odot}$ and 0.9~M$_{\odot}$ carbon-oxygen white dwarf might account for some highly polarised objects such as SN~2004dt but not for the typical behaviour observed in SNe~Ia. In particular, our study consistently finds that the discrepancies can be attributed to the location and nature of the ashes of the companion star. 
%The results presented here, however, do not rule out the violent merger model as a whole since the ejecta morphology for other merger scenarios can be substantially different \citep{pakmor2010,pakmor2013,kromer2013}. 
Future studies will focus on studying polarisation signatures for other existing merger models \citep[e.g.][]{pakmor2013,kromer2013} and on investigating alternative scenarios with revised explosion/secondary properties (e.g. different mass ratios, a different location and time of ignition or a larger distance between the two stars). This might reduce the ejecta asymmetries and thus lead to better agreement with spectropolarimetric data of SNe~Ia.

% These two optical features have been identified in some SNe Ia spectra and suggested as a discriminator among different explosion models \citep{parrent2011,thomas2011}. 

%In the last few decades, spectropolarimetric studies of SNe~Ia have shown that the degree of polarisation in these objects is typically low and tends to decrease after maximum light, indicating that the outer layers are more asymmetric than the inner layers \citep[see][for a review]{wang2008}. However, understanding the exact temporal behaviour is challenging because of the shortage of multi-epoch spectropolarimetric data. In Section~\ref{evolution} we investigated the spectral evolution of the violent merger model of \citet{pakmor2012} for 35 different observer orientations. We found the evolution of about half of the viewing angles selected (including $\bmath{n_1}$) to clearly deviate from the observed trend, with polarisation levels that are typically too high and still increasing about 10~days after maximum (see Fig.~\ref{vlcpol} and Fig.~\ref{polvsbrightness}). Percentage levels and temporal behaviour for the remaining viewing angles, instead, are more consistent with observations. 

\section*{Acknowledgements}

%The authors are thankful to Douglas Leonard and Justyn Maund for providing the spectropolarimetric data of SN~2004dt and SN~2012fr reported in this paper. 

The authors are thankful to Douglas Leonard and Justyn Maund, for providing the spectropolarimetric data of SN~2004dt and SN~2012fr reported in this paper, and to Nando Patat, Lifan Wang and Craig Wheeler for useful discussions. 

SAS acknowledges support from STFC grant ST/L000709/1, RP by the European Research Council under ERC-StG 
grant EXAGAL-308037, and WH and ST by TRR~33 ``The Dark Universe" of the German Research Foundation (DFG). FKR acknowledges support by the DAAD/Go8 German-Australian exchange program, and by the ARCHES prize of the German Ministry of Education and Research (BMBF). IRS was supported by the Australian Research Council Laureate Grant FL0992131.

This work used the DiRAC Complexity system, operated by the University of Leicester IT Services, which forms part of the STFC DiRAC HPC Facility (www.dirac.ac.uk). This equipment is funded by BIS National E-Infrastructure capital grant ST/K000373/1 and  STFC DiRAC Operations grant ST/K0003259/1. DiRAC is part of the National E-Infrastructure.

This research was supported by the Partner Time Allocation (Australian National University), the National Computational Merit Allocation and the Flagship Allocation Schemes of the NCI National Facility at the Australian National University. Parts of this research were conducted by the Australian Research Council Centre of Excellence for All-sky Astrophysics (CAASTRO), through project number CE110001020.

\bibliographystyle{mn2e}
\bibliography{bulla2015b}

\begin{thebibliography}{}

\bibitem[\protect\citeauthoryear{Benz, Cameron, Press \& Bowers}{Benz
  et~al.}{1990}]{benz1990}
Benz W.,  Cameron A.,  Press W.,    Bowers R.,  1990, ApJ, 348, 647

\bibitem[\protect\citeauthoryear{Bulla, Sim \& Kromer}{Bulla
  et~al.}{2015}]{bulla2015}
Bulla M.,  Sim S.~A.,    Kromer M.,  2015, MNRAS, 450, 967

\bibitem[\protect\citeauthoryear{Dan, Rosswog, Brueggen \& Podsiadlowski}{Dan
  et~al.}{2014}]{dan2014}
Dan M.,  Rosswog S.,  Brueggen M.,    Podsiadlowski P.,  2014, MNRAS, 438, 14

\bibitem[\protect\citeauthoryear{Dessart, Hillier, Blondin \& Khokhlov}{Dessart
  et~al.}{2014}]{dessart2014}
Dessart L.,  Hillier D.~J.,  Blondin S.,    Khokhlov A.,  2014, MNRAS, 441,
  3249

\bibitem[\protect\citeauthoryear{Gall, Taubenberger, Kromer, Sim, Benetti,
  Blanc, Elias-Rosa \& Hillebrandt}{Gall et~al.}{2012}]{gall2012}
Gall E.,  Taubenberger S.,  Kromer M.,  Sim S.,  Benetti S.,  Blanc G.,
  Elias-Rosa N.,    Hillebrandt W.,  2012, MNRAS, 427, 994

\bibitem[\protect\citeauthoryear{Guillochon, Dan, Ramirez-Ruiz \&
  Rosswog}{Guillochon et~al.}{2010}]{guillochon2010}
Guillochon J.,  Dan M.,  Ramirez-Ruiz E.,    Rosswog S.,  2010, ApJL, 709, L64

\bibitem[\protect\citeauthoryear{Hillebrandt, Kromer, R{\"o}pke \&
  Ruiter}{Hillebrandt et~al.}{2013}]{hillebrandt2013}
Hillebrandt W.,  Kromer M.,  R{\"o}pke F.,    Ruiter A.,  2013, Frontiers of
  Physics, 8, 116

\bibitem[\protect\citeauthoryear{Iben \& Tutukov}{Iben \&
  Tutukov}{1984}]{iben1984}
Iben J.~I.,  Tutukov A.~V.,  1984, ApJS, 54, 335

\bibitem[\protect\citeauthoryear{Kasen, Nugent, Thomas \& Wang}{Kasen
  et~al.}{2004}]{kasen2004}
Kasen D.,  Nugent P.,  Thomas R.~C.,    Wang L.,  2004, ApJ, 610, 876

\bibitem[\protect\citeauthoryear{Kashyap, Fisher, Garc{\'\i}a-Berro,
  Aznar-Sigu{\'a}n, Ji \& Lor{\'e}n-Aguilar}{Kashyap
  et~al.}{2015}]{kashyap2015}
Kashyap R.,  Fisher R.,  Garc{\'\i}a-Berro E.,  Aznar-Sigu{\'a}n G.,  Ji S.,
  Lor{\'e}n-Aguilar P.,  2015, ApJL, 800, L7

\bibitem[\protect\citeauthoryear{Kromer, Pakmor, Taubenberger, Pignata, Fink,
  R{\"o}pke, Seitenzahl, Sim \& Hillebrandt}{Kromer et~al.}{2013}]{kromer2013}
Kromer M.,  Pakmor R.,  Taubenberger S.,  Pignata G.,  Fink M.,  R{\"o}pke F.,
  Seitenzahl I.,  Sim S.,    Hillebrandt W.,  2013, ApJL, 778, L18

\bibitem[\protect\citeauthoryear{Kromer \& Sim}{Kromer \&
  Sim}{2009}]{kromer2009}
Kromer M.,  Sim S.~A.,  2009, MNRAS, 398, 1809

\bibitem[\protect\citeauthoryear{Leonard, Li, Filippenko, Foley \&
  Chornock}{Leonard et~al.}{2005}]{leonard2005}
Leonard D.~C.,  Li W.,  Filippenko A.~V.,  Foley R.~J.,    Chornock R.,  2005,
  ApJ, 632, 450

\bibitem[\protect\citeauthoryear{Maoz, Mannucci \& Nelemans}{Maoz
  et~al.}{2014}]{maoz2014}
Maoz D.,  Mannucci F.,    Nelemans G.,  2014, ARA\&A, 52

\bibitem[\protect\citeauthoryear{Maund, Spyromilio, H{\"o}flich, Wheeler,
  Baade, Clocchiatti, Patat, Reilly, Wang \& Zelaya}{Maund
  et~al.}{2013}]{maund2013}
Maund J.~R.,  Spyromilio J.,  H{\"o}flich P.,  Wheeler J.,  Baade D.,
  Clocchiatti A.,  Patat F.,  Reilly E.,  Wang L.,    Zelaya P.,  2013, MNRAS,
  433, L20

\bibitem[\protect\citeauthoryear{Moll, Raskin, Kasen \& Woosley}{Moll
  et~al.}{2014}]{moll2014}
Moll R.,  Raskin C.,  Kasen D.,    Woosley S.~E.,  2014, ApJ, 785, 105

\bibitem[\protect\citeauthoryear{Pakmor, Kromer, R{\"o}pke, Sim, Ruiter \&
  Hillebrandt}{Pakmor et~al.}{2010}]{pakmor2010}
Pakmor R.,  Kromer M.,  R{\"o}pke F.~K.,  Sim S.~A.,  Ruiter A.~J.,
  Hillebrandt W.,  2010, Nature, 463, 61

\bibitem[\protect\citeauthoryear{Pakmor, Kromer, Taubenberger, Sim, R{\"o}pke
  \& Hillebrandt}{Pakmor et~al.}{2012}]{pakmor2012}
Pakmor R.,  Kromer M.,  Taubenberger S.,  Sim S.~A.,  R{\"o}pke F.~K.,
  Hillebrandt W.,  2012, ApJL, 747, L10

\bibitem[\protect\citeauthoryear{Pakmor, Kromer, Taubenberger \&
  Springel}{Pakmor et~al.}{2013}]{pakmor2013}
Pakmor R.,  Kromer M.,  Taubenberger S.,    Springel V.,  2013, ApJL, 770, L8

\bibitem[\protect\citeauthoryear{Parrent, Thomas, Fesen, Marion, Challis,
  Garnavich, Milisavljevic, Vink{\`o} \& Wheeler}{Parrent
  et~al.}{2011}]{parrent2011}
Parrent J.~T.,  Thomas R.~C.,  Fesen R.~A.,  Marion G.~H.,  Challis P.,
  Garnavich P.~M.,  Milisavljevic D.,  Vink{\`o} J.,    Wheeler J.~C.,  2011,
  ApJ, 732, 30

\bibitem[\protect\citeauthoryear{Patat, Baade, H{\"o}flich, Maund, Wang \&
  Wheeler}{Patat et~al.}{2009}]{patat2009}
Patat F.,  Baade D.,  H{\"o}flich P.,  Maund J.~R.,  Wang L.,    Wheeler J.~C.,
   2009, A\&A, 508, 229

\bibitem[\protect\citeauthoryear{Phillips, Lira, Suntzeff, Schommer, Hamuy \&
  Maza}{Phillips et~al.}{1999}]{phillips1999}
Phillips M.,  Lira P.,  Suntzeff N.~B.,  Schommer R.,  Hamuy M.,    Maza J.,
  1999, AJ, 118, 1766

\bibitem[\protect\citeauthoryear{Phillips}{Phillips}{1993}]{phillips1993}
Phillips M.~M.,  1993, ApJ, 413, L105

\bibitem[\protect\citeauthoryear{Raskin, Kasen, Moll, Schwab \& Woosley}{Raskin
  et~al.}{2014}]{raskin2014}
Raskin C.,  Kasen D.,  Moll R.,  Schwab J.,    Woosley S.,  2014, ApJ, 788, 75

\bibitem[\protect\citeauthoryear{R{\"o}pke, Seitenzahl, Benitez, Fink, Pakmor,
  Kromer, Sim, Ciaraldi-Schoolmann \& Hillebrandt}{R{\"o}pke
  et~al.}{2011}]{roepke2011}
R{\"o}pke F.,  Seitenzahl I.,  Benitez S.,  Fink M.,  Pakmor R.,  Kromer M.,
  Sim S.,  Ciaraldi-Schoolmann F.,    Hillebrandt W.,  2011, Progress in
  Particle and Nuclear Physics, 66, 309

\bibitem[\protect\citeauthoryear{R{\"o}pke, Kromer, Seitenzahl, Pakmor, Sim,
  Taubenberger, Ciaraldi-Schoolmann, Hillebrandt, Aldering, Antilogus
  et~al.,}{R{\"o}pke et~al.}{2012}]{roepke2012}
R{\"o}pke F.~K.,  Kromer M.,  Seitenzahl I.~R.,  Pakmor R.,  Sim S.~A.,
  Taubenberger S.,  Ciaraldi-Schoolmann F.,  Hillebrandt W.,  Aldering G.,
  Antilogus P.,    et~al., 2012, ApJL, 750, L19

\bibitem[\protect\citeauthoryear{Scalzo, Aldering, Antilogus, Aragon, Bailey,
  Baltay, Bongard, Buton, Cellier-Holzem, Childress et~al.,}{Scalzo
  et~al.}{2014}]{scalzo2014}
Scalzo R.,  Aldering G.,  Antilogus P.,  Aragon C.,  Bailey S.,  Baltay C.,
  Bongard S.,  Buton C.,  Cellier-Holzem F.,  Childress M.,    et~al., 2014,
  MNRAS, 440, 1498

\bibitem[\protect\citeauthoryear{Seitenzahl, Meakin, Townsley, Lamb \&
  Truran}{Seitenzahl et~al.}{2009}]{seitenzahl2009}
Seitenzahl I.~R.,  Meakin C.~A.,  Townsley D.~M.,  Lamb D.~Q.,    Truran J.~W.,
   2009, ApJ, 696, 515

\bibitem[\protect\citeauthoryear{Seitenzahl, Summa, Krau{\ss}, Sim, Diehl,
  Els{\"a}sser, Fink, Hillebrandt, Kromer, Maeda et~al.,}{Seitenzahl
  et~al.}{2015}]{seitenzahl2015}
Seitenzahl I.~R.,  Summa A.,  Krau{\ss} F.,  Sim S.~A.,  Diehl R.,
  Els{\"a}sser D.,  Fink M.,  Hillebrandt W.,  Kromer M.,  Maeda K.,    et~al.,
  2015, MNRAS, 447, 1484

\bibitem[\protect\citeauthoryear{Shen, Bildsten, Kasen \& Quataert}{Shen
  et~al.}{2012}]{shen2012}
Shen K.~J.,  Bildsten L.,  Kasen D.,    Quataert E.,  2012, ApJ, 748, 35

\bibitem[\protect\citeauthoryear{Sim}{Sim}{2007}]{sim2007}
Sim S.~A.,  2007, MNRAS, 375, 154

\bibitem[\protect\citeauthoryear{Summa, Ulyanov, Kromer, Boyer, R{\"o}pke, Sim,
  Seitenzahl, Fink, Mannheim, Pakmor et~al.,}{Summa et~al.}{2013}]{summa2013}
Summa A.,  Ulyanov A.,  Kromer M.,  Boyer S.,  R{\"o}pke F.~K.,  Sim S.~A.,
  Seitenzahl I.~R.,  Fink M.,  Mannheim K.,  Pakmor R.,    et~al., 2013, A\&A,
  554, A67

\bibitem[\protect\citeauthoryear{Thomas, Aldering, Antilogus, Aragon, Bailey,
  Baltay, Bongard, Buton, Canto, Childress et~al.,}{Thomas
  et~al.}{2011}]{thomas2011}
Thomas R.~C.,  Aldering G.,  Antilogus P.,  Aragon C.,  Bailey S.,  Baltay C.,
  Bongard S.,  Buton C.,  Canto A.,  Childress M.,    et~al., 2011, ApJ, 743,
  27

\bibitem[\protect\citeauthoryear{Van~Kerkwijk, Chang \& Justham}{Van~Kerkwijk
  et~al.}{2010}]{vankerkwijk2010}
Van~Kerkwijk M.~H.,  Chang P.,    Justham S.,  2010, The Astrophysical Journal
  Letters, 722, L157

\bibitem[\protect\citeauthoryear{Wang, Baade, H{\"o}flich, Wheeler, Kawabata,
  Khokhlov, Nomoto \& Patat}{Wang et~al.}{2006}]{wang2006}
Wang L.,  Baade D.,  H{\"o}flich P.,  Wheeler J.~C.,  Kawabata K.,  Khokhlov
  A.,  Nomoto K.,    Patat F.,  2006, ApJ, 653, 490

\bibitem[\protect\citeauthoryear{Wang, Baade \& Patat}{Wang
  et~al.}{2007}]{wang2007}
Wang L.,  Baade D.,    Patat F.,  2007, Science, 315, 212

\bibitem[\protect\citeauthoryear{Wang \& Wheeler}{Wang \&
  Wheeler}{2008}]{wang2008}
Wang L.,  Wheeler J.~C.,  2008, ARA\&A, 46, 433

\bibitem[\protect\citeauthoryear{Webbink}{Webbink}{1984}]{webbink1984}
Webbink R.~F.,  1984, ApJ, 277, 355

\bibitem[\protect\citeauthoryear{Whelan \& Iben}{Whelan \&
  Iben}{1973}]{whelan1973}
Whelan J.,  Iben J.~I.,  1973, ApJ, 186, 1007

\bibitem[\protect\citeauthoryear{Zhu, Chang, van Kerkwijk \& Wadsley}{Zhu
  et~al.}{2013}]{zhu2013}
Zhu C.,  Chang P.,  van Kerkwijk M.~H.,    Wadsley J.,  2013, ApJ, 767, 164

\end{thebibliography}

\end{document}